\documentclass[numberedappendix]{emulateapj}
\usepackage{amsmath,amssymb,graphicx,natbib}

\renewcommand{\d}{\partial}
\newcommand{\const}{\mathrm{const}}
\newcommand{\Mbh}{M_\bullet}
\newcommand{\CoefN}{Q}
\newcommand{\CoefK}{k}
\newcommand{\CoefA}{\mathcal A}
\newcommand{\CoefLn}{b}
\newcommand{\epsilond}{\epsilon_d}
\newcommand{\epsilonp}{\epsilon_p}
\newcommand{\Ham}{\tilde H}
\newcommand{\calH}{\mathcal H}
\newcommand{\calR}{\mathcal R}
\newcommand{\calRmin}{{\mathcal R}_\mathrm{min}}
\newcommand{\calRmax}{{\mathcal R}_\mathrm{max}}
\newcommand{\calRlow}{{\mathcal R}_\mathrm{low}}
\newcommand{\calRsep}{\mathcal R_\mathrm{sep}}
\newcommand{\calRcapt}{\mathcal R_\mathrm{lc}}
\newcommand{\calReff}{\mathcal R_\mathrm{eff}}

\newcommand{\calRch}{\mathcal R_\mathrm{ch}}
\newcommand{\calG}{\mathcal G}
\newcommand{\calGav}{\mathcal G_\mathrm{av}}
\newcommand{\calGsph}{\mathcal G_\mathrm{sph}}
\newcommand{\calE}{\mathcal E}
\newcommand{\calQ}{\mathcal Q}
\newcommand{\calF}{\mathcal F}
\newcommand{\calN}{\mathcal N}
\newcommand{\calFdrain}{\calF_\mathrm{drain}}
\newcommand{\Adv}{\mathcal D}
\newcommand{\Dif}{\mathcal D}
\newcommand{\dvpar}{\langle \Delta v_\| \rangle}
\newcommand{\dvsqpar}{\langle \Delta v^2_\| \rangle}
\newcommand{\dvsqper}{\langle \Delta v^2_\bot \rangle}
\newcommand{\Lsqinv}{\kappa\,}

\newcommand{\Lpp}{\frac{\Lsqinv''}{\Lsqinv}}
\newcommand{\Nfull}{N_\mathrm{full\ lc}}
\newcommand{\Nreal}{N_\mathrm{real\ lc}}

\newcommand{\trad}{T_\mathrm{rad}}
\newcommand{\tprec}{T_\mathrm{M}}
\newcommand{\tosc}{T_\mathrm{prec}}
\newcommand{\trel}{T_\mathrm{rel}}
\newcommand{\rh}{r_\mathrm{m}}
\newcommand{\rtid}{r_\mathrm{tid}}

\newcommand{\rg}{r_\mathrm{g}}
\newcommand{\Lcirc}{L_\mathrm{circ}}
\newcommand{\sbh}{SBH}
\newcommand{\sbhs}{SBHs}
\begin{document}

\title{The loss cone problem in axisymmetric nuclei}

\author{Eugene Vasiliev}
\affil{School of Physics and Astronomy and Center for Computational Relativity and Gravitation, \protect\\
Rochester Institute of Technology, Rochester, NY, USA \protect\\
and Lebedev Physical Institute, Moscow, Russia}
\email{eugvas@lpi.ru}
\author{David Merritt}
\affil{School of Physics and Astronomy and Center for Computational Relativity and Gravitation, \protect\\
Rochester Institute of Technology, Rochester, NY, USA}
\email{merritt@astro.rit.edu}

\begin{abstract}
We consider the problem of consumption of stars by
a supermassive black hole (\sbh) at the center of an axisymmetric galaxy.
Inside the \sbh\ sphere of influence, motion of stars in the mean field
is regular and can be described analytically in terms of three integrals of motion:
the energy $E$, the $z$-component of angular momentum $L_z$,
and the secular Hamiltonian $H$.
There exist two classes of orbits, tubes and saucers; saucers occupy the 
low-angular-momentum parts of phase space and their fraction is proportional to the
degree of flattening of the nucleus.
Perturbations due to gravitational encounters lead to diffusion of stars
in integral space, which can be described using the Fokker-Planck equation.
We calculate the diffusion coefficients and solve this equation in the two-dimensional
phase space ($L_z, H$), for various values of the capture radius and the degree of flattening.
Capture rates are found to be modestly higher than in the spherical case,
up to a factor of a few,
and most captures take place from saucer orbits.
We also carry out a set of collisional $N$-body simulations to confirm the predictions
of the Fokker-Planck models.
We discuss the implications of our results for  rates of tidal disruption and capture
in the Milky Way and external galaxies.
\end{abstract}

\section{Introduction}

The study of collisional relaxation in stellar nuclei around massive black holes
and the associated rates of capture has a long history.
The pioneering work of \citet{BahcallWolf1976} established
a quasi-steady-state solution for the stellar distribution,
now known as a Bahcall-Wolf cusp, which has a density $\rho(r) \propto r^{-7/4}$
inside the radius of influence $\rh$, defined roughly as the radius enclosing
a mass in stars equal to the mass $M_\bullet$ of the hole.
Their solution was obtained from the steady-state,
one-dimensional Fokker-Planck equation
describing two-body relaxation and energy exchange between stars in the
(Newtonian) gravitational field of the massive object, and is characterized
by zero (or very small) flux of stars with respect to energy into the central hole.

A more refined treatment requires the concept of a ``loss cone'', the region of
phase space corresponding to stars with sufficienly low angular momenta
to be captured by the black hole: $L<L_\mathrm{lc}$, where the capture boundary
$L_\mathrm{lc}\approx\sqrt{2G\Mbh r_\mathrm{lc}}$ is determined either by 
tidal disruption or by direct capture, at some radius $r_\mathrm{lc}$
\citep{FrankRees1976, LightmanShapiro1977}.
The latter paper also introduced the important distinction between empty- and
full-loss-cone regimes, with the boundary between them defined as the energy
at which the typical change of angular momentum in one radial period,
$\sqrt{\langle\Delta L^2\rangle|_{\trad}}$, is equal to $L_\mathrm{lc}$.
\citet{LightmanShapiro1977} derived the quasi-steady-state rate of
consumption of stars as a function of energy and showed that the distribution
function depends logarithmically on angular momentum near the loss cone.
These authors, and subsequently \citet{BahcallWolf1977}, included capture from the
loss cone via an energy-dependent sink term in the one-dimensional Fokker-Planck
equation for $f(E,t)$.
While the addition of such a term greatly increases the capture rate, it was found
to have little effect on the form of the density profile at radii
$r_\mathrm{lc} \ll r \lesssim \rh$.
\citet{CohnKulsrud1978} solved two-dimensional
Fokker-Planck equation in $(E,L)$ space and confirmed the results of the more
approximate one-dimensional studies.

These early studies were targeted toward massive black holes in globular clusters.
The theory was subsequently applied to determine capture rates in galactic nuclei 
\citep{SyerUlmer1999, MagorrianTremaine1999, WangMerritt2004, SirotaIZG2005}.
Studies based on the Fokker-Planck equation were also verified by other methods
such as Monte-Carlo models \citep{ShapiroMarchant1978, FreitagBenz2002},
gaseous models \citep{AmaroSeoaneFS2004} and direct $N$-body simulation
\citep{BaumgardtME2004, PretoMS2004, KomossaMerritt2008,BrockampBK2011}.

Relaxation times in galactic nuclei are much longer than in globular clusters,
and in many cases much longer than galaxy lifetimes.
Partly as a consequence, galactic nuclei need not be spherically symmetric,
and they may contain a substantial population of ``centrophilic'' orbits
(saucers, pyramids) that dominate the capture rate, even in the absence
of gravitational encounters \citep{NormanSilk1983, GerhardBinney1985, 
MagorrianTremaine1999, MerrittPoon2004, MerrittVasiliev2011}.
In the context of collisional loss-cone repopulation,
by far the majority of studies have assumed spherical symmetry.
This assumption is not crucial for $N$-body integrations (except insofar as it can be
difficult to construct nonspherical initial conditions), but it is an important ingredient of
Fokker-Planck studies, because it ensures the conservation of angular momentum
in the unperturbed motion.
To date, all Fokker-Planck treatments that allowed for non-sphericity
\citep{Goodman1983,EinselSpurzem1999, FiestasSK2006, FiestasSpurzem2010}
have assumed axisymmetry and have further restricted the allowed form of $f$
by writing $f=f(E,L_z)$, with $L_z$ the component of angular momentum parallel
to the symmetry axis.
Two integrals of motion are not sufficient to specify regular motion in the
axisymmetric geometry however,
and numerical integrations in axisymmetric potentials
typically reveal a third integral, $I_3$ (with some orbits chaotic).
Of course, the reason for the neglect of $I_3$ is the
absence of knowledge about its functional form.
For mildly flattened systems, $I_3$ can be approximated by $L^2$ \citep{Saaf1968},
and this approximation has been used as a basis
for constructing steady-state models \citep[e.g.][]{LuptonGunn1987}.

The neglect of the third integral in the axisymmetric problem
has several important consequences.
Instead of individual orbits populated by stars having the same values of
their integrals of motion, one effectively considers ensembles of orbits
composed of stars with different $I_3$.
Moreover, by setting $f=f(E,L_z)$, the phase space density is forced 
to be uniform within this ensemble,
which may lead to unphysical constraints on the possible evolution of the system.
The diffusion coefficients must also be evaluated as if they did not
depend on $I_3$, or, more correctly, are averaged over all possible values of
$I_3$.
Finally, ignoring the third integral precludes the detailed study of regular
orbits such as saucers, which might be expected to dominate the loss rate 
\citep{MagorrianTremaine1999}.
However, sufficiently close to the black hole, the unperturbed motion is
nearly Keplerian, and standard ``planetary'' perturbation theory implies
the existence of a third integral, which can sometimes even be expressed
in terms of simple functions \citep{SridharTouma1999}.
In this approximation, the unperturbed stellar orbits are regular (integrable) and respect
three independent integrals of the motion: $E$, $L_z$ and $H$, where
$H$ is the ``secular,'' i.e. averaged, Hamiltonian.

Given an analytic expression for the third integral in the vicinity of the black hole,
fully three-dimensional Fokker-Planck studies become feasible, describing the
time evolution of $f=f(E,L_z,H)$.
For the present study, however, we choose to concentrate on evolution in
the two-dimensional subspace $(L_z, H)$ with the gravitational potential,
and the orbital energy $E$, fixed.
Our justification for ignoring changes in $E$ is the same as in many previous
studies of the loss-cone problem in galactic nuclei \citep{SyerUlmer1999,MagorrianTremaine1999}:
energy relaxation time scales are typically very long in nuclei, too long for
steady-state configurations like the Bahcall-Wolf cusp to be reached.
Instead, the dependence of $f$ on $E$ is inferred from the observed, radial
density profile via Eddington's formula.
Our goal is to generalize the well-known, one-dimensional solutions
for $f(L)$ in the spherical geometry to the two-dimensional case
$f(L_z,H)$ in the axisymmetric geometry.

The paper is organized as follows.
In \S\ref{sec_orbital_structure} we use the averaging method to demonstrate
the existence of a third integral of motion inside the supermassive black hole
(\sbh) sphere of influence and we use it to elucidate the behavior of orbits:
the tube orbits that are generic to the axisymmetric geometry,
and the saucer orbits that
inhabit the low-angular-momentum parts of phase space.
The Fokker-Planck equation is written down in \S\ref{sec_FP_definition},
and a scheme for calculating diffusion coefficients is presented in the case
of three integrals of motion.
Following this general treatment, we then restrict our attention to weakly
flattened systems, which allows some simplification in the computations.
We also concentrate on the case $\rho(r)\propto r^{-3/2}$, which is both
physically reasonable, and which results in analytic expressions for
many of the diffusion coefficients.
\S\ref{sec_boundary_cond} discusses the proper boundary conditions for the
Fokker-Planck equation, and \S\ref{sec_FP_solution} is devoted to the
solution of the two-dimensional equation and comparison
with the one-dimensional (spherical) case.
It turns out that the flux of stars into the \sbh\ is enhanced with
respect to the spherical case, but by a modest factor: at most a factor of a few.
In \S\ref{sec_Nbody} we describe direct $N$-body simulations designed to
test the predictions of the Fokker-Planck models; 
sections~\ref{sec_chaos} and \ref{sec_triaxiality} briefly discuss the role of
chaotic orbits beyond the \sbh\ influence sphere, and triaxiality of the stellar
potential, respectively.
Finally, in \S\ref{sec_estimates} we estimate capture rates in realistic 
galaxy models, using the Fokker-Planck models to access the range of
parameters not presently accessible to $N$-body simulations.

\section{Motion in axisymmetric star clusters around black holes}
\label{sec_orbital_structure}

Consider a stellar nucleus in which the density varies as a power of radius, 
 $\rho\sim r^{-\gamma}$, 
and in which the equidensity contours are flattened in the direction of the
short ($z$) axis; in other words, an oblate system.
Let $p\le 1$ be the axis ratio, i.e. the ratio of radii along the minor and 
major axes at which densities are equal.
In the first approximation, the stellar density and potential of a flattened 
system are described by the spherical part modified by an $l=2$ Legendre polynomial:
\begin{subequations}\label{eq_modelrhophi}
\begin{eqnarray} 
\rho_\star(\boldsymbol{x}) &=& \rho_0 \left(\frac{r}{r_0}\right)^{-\gamma}\, 
  \left(1 + \epsilond \left[\frac{z^2}{r^2}-\frac{1}{3}\right] \right) ,  \label{eq_rho_cusp} \\
\Phi_\star(\boldsymbol{x}) &=& \Phi_0 \left(\frac{r}{r_0}\right)^{2-\gamma}\,
  \left(1 + \epsilonp \left[\frac{z^2}{r^2}-\frac{1}{3}\right] \right) ,  \label{eq_phi_cusp} \\
\Phi_0 &=& \frac{4\pi\,G\rho_0\,r_0^2}{(3-\gamma)(2-\gamma)} \;,  \label{eq_phi0} \\
\epsilonp &=& \epsilond\,\frac{(3-\gamma)(2-\gamma)}{\gamma\,(\gamma-5)} \;,  \label{eq_epsilonp}\\
\epsilond &=& -\frac{3 (p^{-\gamma}-1)}{2p^{-\gamma}+1}   \label{eq_epsilond}
\end{eqnarray}
\end{subequations}
\quad 
where $0\le \gamma < 2$. 
The total gravitational potential is 
\begin{equation}
\Phi(\boldsymbol{x}) = -\frac{G\Mbh}{r} + \Phi_\star(\boldsymbol{x}) 
\end{equation}
where $\Mbh$ is the mass of the supermassive black hole (\sbh) located
at $\boldsymbol{x}=0$.

Throughout this section, we restrict attention to motion that satisfies
\begin{equation}  \label{eq_rinfl}
\rg \equiv \frac{G\Mbh}{c^2} \ll r \ll \rh \equiv 
  r_0 \left[ \frac{\Mbh\,(3-\gamma)}{2\pi\rho_0 r_0^3} \right]^{1/(3-\gamma)} ,
\end{equation}
where $\rg$ is the gravitational radius of the \sbh\
and $\rh$ its radius of  influence; the latter is conventionally defined as the
radius of a sphere containing a mass in stars equal to $2\Mbh$.
The first inequality permits us to ignore relativistic corrections to the equations 
of motion, and the second allows us to treat the force from the stars
as a small perturbation to the inverse-square force from the \sbh.
The effects of relativity are discussed briefly in Appendix~\ref{sec_appendix_relativity},
where more precise conditions for the validity of the Newtonian approximation are
derived.

Under these conditions, one expects the motion to be nearly Keplerian on time
scales comparable with the radial period, and we can employ the method of averaging
\citep{SridharTouma1999, SambhusSridhar2000}:
the forces acting on a star are averaged over the unperturbed motion, with the 
orbital elements -- the ``osculating elements'' --  fixed during the averaging. 
It is convenient to describe the motion using the Delaunay variables which are
action-angle variables in the unperturbed problem:
$I, L, L_z$ (actions) and $w, \omega, \Omega$ (angles). Here 
\begin{equation}  \label{eq_Lcirc}
I = \Lcirc \equiv \sqrt{G\Mbh a} = \frac{G\Mbh}{\sqrt{-2E}}
\end{equation} 
is the angular momentum of a circular orbit
 with given semimajor axis $a$ or total energy $E$, 
$L$ is the magnitude and $L_z$ is the $z-$component of the angular momentum 
(so that $\cos i\equiv L_z/L$ gives the inclination of orbital plane with respect 
to the $x-y$ plane);
$w$ is the radial phase (mean anomaly), 
$\omega$ is the argument of periastron ($\omega=0$ corresponds to periapsis
in $x-y$ plane), and $\Omega$ is the longitude of  ascending node.
We further define dimensionless angular momentum variables as $\ell\equiv L/I \in[0,1]$ 
and $\ell_z\equiv L_z/I \in[-\ell,\ell]$.
We will also have occasion to refer to their squared values, denoted, following 
\citet{CohnKulsrud1978}, as $\calR\equiv \ell^2$ and  $\calR_z\equiv \ell_z^2$.

These three pairs of canonically-conjugate variables evolve according to Hamilton's 
equations of motion, with the Hamiltonian given by
\begin{equation}
H = -\frac{1}{2}\left(\frac{G\Mbh}{I}\right)^2 + \Phi_\star 
\end{equation}
and with $\Phi_\star$ expressed in terms of the Delaunay variables.
We assume that these variables -- with the exception of the radial phase $w$ -- 
are nearly constant over one radial period:
\begin{equation}  \label{eq_Trad}
\trad \equiv \frac{2\pi a^{3/2}}{\sqrt{G\Mbh}}.
\end{equation}
The averaged equations of motion can then be defined as the equations of motion 
corresponding to the averaged  Hamiltonian
\begin{eqnarray}
\overline H &\equiv& \frac{1}{2\pi} \oint dw\,H \\
&=& -\frac{1}{2}\left(\frac{G\Mbh}{I}\right)^2 + 
\frac{1}{2\pi}\oint \Phi_\star\left(I,L,L_z,\omega,\Omega;w\right) dw \nonumber 
\end{eqnarray}
where  the variables $\{I,L,L_z,\omega,\Omega\}$ are fixed 
in the averaging of $\Phi_\star$ over $w$.

After the averaging, $\overline H$ is independent of $w$ and therefore $I$ is conserved, 
as is the semimajor axis $a$ and the energy $E$.
On the other hand, $\overline H$ itself is a (new) integral of the motion.
Finally, $L_z$ is conserved due to axial symmetry, from which it follows that the motion 
is integrable. Remarkably, even in the (weakly) triaxial case there can exist three integrals 
of motion, $L_z$ being replaced by another conserved quantity \citep{MerrittVasiliev2011}.

Exact expressions for the averaged Hamiltonian are given in Appendix~\ref{sec_appendix_Hamiltonian}.
A good approximation to the averaged perturbing potential is
\begin{subequations}\label{eq_Happrox}
\begin{eqnarray}
\overline \Phi_\star &=& \Phi_0\left(\frac{a}{r_0}\right)^{2-\gamma}\, \Ham(\ell,\ell_z, \omega, \Omega) ,\\
\Ham &\approx& \left(1-\epsilonp/3\right) (\CoefN - (\CoefN-1)\ell^2) +  \label{eq_Hamiltonian_approx} \\
 &+& \epsilonp\left(1-\frac{\ell_z^2}{\ell^2}\right)
  \left( \CoefN(1-\ell^2)\sin^2\omega + \frac{1}{2} \ell^2 \right) ,\nonumber\\
\CoefN &\equiv& \frac{2^{3-\gamma} \,
  \Gamma\left(\frac{7}{2}-\gamma\right)}{\sqrt{\pi}\,\Gamma(4-\gamma)} ,
\end{eqnarray}
\end{subequations}
which is exact for $\gamma=\{0,1\}$ (for which $\CoefN=5/2,3/2$) and approximates 
the true value to within a few percent in other cases.
This Hamiltonian is very similar to the averaged Hamiltonian of the hierarchical 
restricted three-body problem \citep{Kozai1962, Lidov1962}; 
a detailed comparison is presented in Appendix~\ref{sec_appendix_kozai}.
 
Expressed in terms of a dimensionless time $\tau\equiv 2\pi t/\tprec$,
the equations of motion read
\begin{equation}  \label{eq_motion}
\frac{d\ell}{d\tau} = -\frac{\d\Ham}{\d\omega} \;, \quad
\frac{d\omega}{d\tau} = \frac{\d\Ham}{\d\ell}.
\end{equation}
The equations for $\ell_z$ and $\Omega$ are not needed because 
$\ell_z$ is conserved and because nothing important depends on $\Omega$. 
The time $\tprec$ is%
\footnote{Note that $\tprec$ differs by a factor $2(\CoefN-1)/3$ from $T_\mathrm{prec}$ 
defined in \citet{MerrittVasiliev2011}.}
\begin{equation}  \label{eq_Tprec}
\tprec\equiv \frac{2\pi I}{\Phi_0}\left(\frac{r_0}{a}\right)^{2-\gamma} 
= (2-\gamma)\frac{M_{\bullet}}{M_\star(a)}\trad 
\end{equation}
where $M_\star(r) \equiv 4\pi r^3\rho_\star(r)/(3-\gamma)$ is approximately
the mass in stars within radius $r$.
$\tprec$ is the time associated with precession of the argument of periastron
due to the spherically-distributed mass (the ``mass-precession time'').

\begin{figure*}
$$\includegraphics[angle=-90]{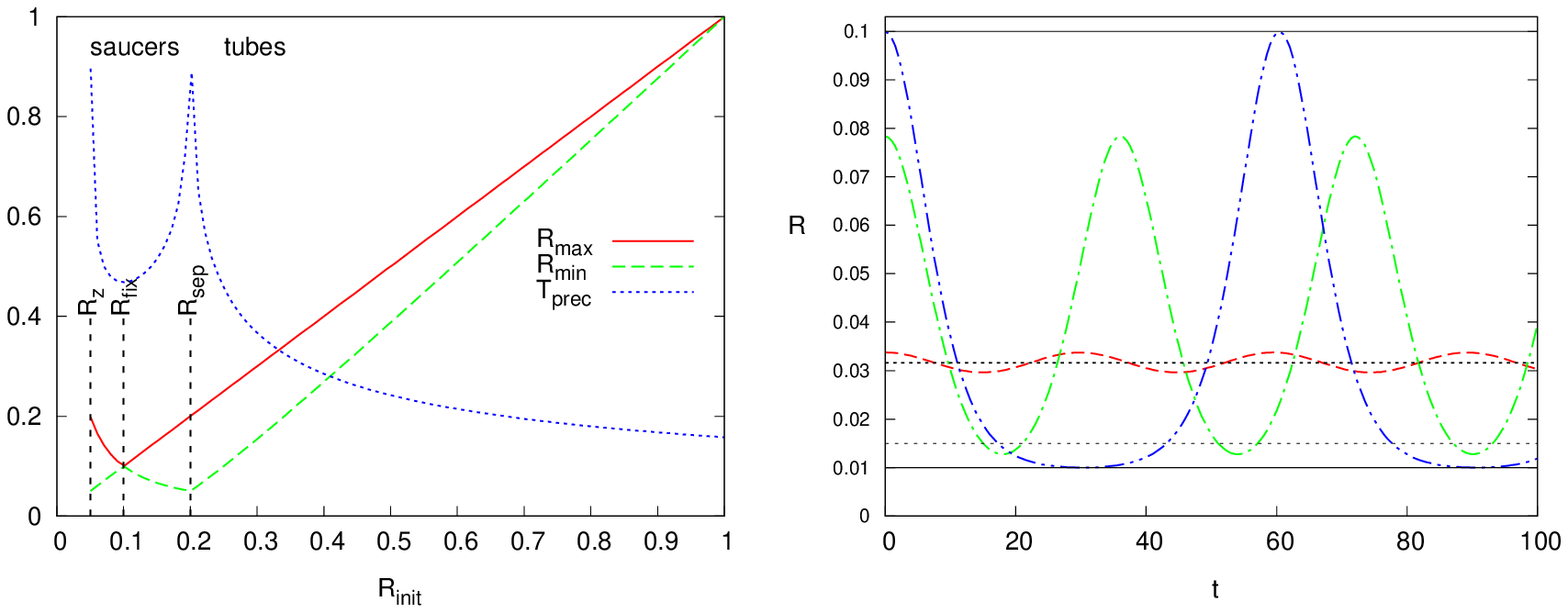}$$
\caption{Left: Maximum and minimum possible values of $\calR$ (solid and dashed lines, 
equation~\ref{eq_Rminmax}) and precession period $\tosc$ (dotted line, equation~\ref{eq_Tosc}) 
for a series of orbits started with $\omega=\pi/2$ and $\calR=\calR_\mathrm{init}$, 
for $\calRsep=0.2$ and $\calR_z=0.05$. Orbits with $\calR_\mathrm{init}<\calRsep$ are saucers, 
others are tubes. Saucer orbits with initial angular momenta $\calR_\mathrm{init}$ and 
$\calR_z\calRsep/\calR_\mathrm{init}$ are identical 
(symmetric about $\calR_\mathrm{fix}=\sqrt{\calRsep\calR_z}$ line). \protect\\
Right: $\calR(t)$ plotted for three saucer orbits (equation~\ref{eq_calR_of_t} 
for $\calRsep=0.1, \calR_z=0.01$).
Red dashed, green dashed-dotted and blue dash-double-dotted lines are for $\calH=-0.048$ 
(close to the fixed-point orbit), $\calH=-0.02$ and $\calH=-0.001$ (close to separatrix).
Thin solid lines show minimum and maximum possible values of $\calR$ for given $\calR_z$ 
(essentially $\calR_z$ and $\calRsep$), and dotted line shows 
$\calR_\mathrm{fix}=\sqrt{\calRsep\calR_z}\approx 0.032$.
Another dotted line shows the capture boundary at $\calRcapt=0.015$.
} \label{fig_Rminmax}
\end{figure*}

In what follows, it will be convenient to replace $\Ham$ by $\calH$, 
a linear combination of $\Ham$ and $\calR_z$: 
\begin{subequations}\label{eq_calH}
\begin{eqnarray}
\calH(\calR,\omega) &\equiv& \frac{\CoefN(1-\epsilonp/3) - \Ham - (\CoefN-1)(1-\epsilonp/3)\calR_z} 
  {(\CoefN-1)(1-\epsilonp/3)-\epsilonp/2} =  \nonumber\\
  &=& (\calR-\calR_z)\left(1 - 
  \frac{\calRsep}{1-\calRsep} \frac{1-\calR}{\calR} \sin^2\omega \right) ,  \label{eq_calHa} \\
\calRsep &\equiv& 
  \frac{\CoefN \epsilonp}{(\CoefN-1)(1-\epsilonp/3)-\epsilonp/2+\CoefN\epsilonp} . \label{eq_CalHb}
\end{eqnarray}
\end{subequations}

To obtain the solution, we express $\omega$ from (\ref{eq_calH}) 
and substitute it into the first of equations (\ref{eq_motion}):
\begin{subequations}
\begin{eqnarray}
\frac{d\calR}{d\tau} &=& -\CoefK\,
  \sqrt{(\calR_1-\calR)(\calR-\calR_2)(\calR-\calR_3)}   \label{eq_dcalRdtau}, \\
\CoefK &\equiv& \frac{4\CoefN\epsilonp\sqrt{1-\calRsep}}{\calRsep}  
  \approx 4(\CoefN-1) \;\mbox{ for }\epsilonp\ll 1 ,  \nonumber\\
\calR_{1,2} &\equiv& \calR_\star \pm \sqrt{\calR_\star^2 - \calRsep\calR_z},  \label{eq_R12}\\
\calR_3 &\equiv& \calH+\calR_z , \nonumber\\
\calR_\star &\equiv& \frac{1}{2}\left[ \calRsep(1+\calR_z) + (1-\calRsep)(\calH+\calR_z)\right]. 
  \nonumber \\
\calRmax &\equiv& \calR_1 , \label{eq_Rminmax} \\
\calRmin &\equiv& \mathrm{max}(\calR_2,\calR_3) , \nonumber \\
\calRlow &\equiv& \mathrm{min}(\calR_2,\calR_3) . \nonumber
\end{eqnarray}
\end{subequations}
It is clear that $\calR$ is allowed to oscillate between $\calRmin$ and $\calRmax$.
We thus have two classes of orbit, depending on the relation between $\calR_2$ and $\calR_3$.
If $\calR_3\ge\calR_2$, which occurs when $\calH\ge0$, the orbit is an ordinary short-axis tube 
(SAT); in the opposite case ($\calH<0$) the orbit is a saucer, with $\calR_3<\calR_2$.
(A similar class of orbits was described by \citet{LeesSchwarzschild1992} for a $\gamma=2$ 
scale-free potential; see Appendix~\ref{sec_appendix_saucers} for more discussion). 
The main distinction is that for saucer orbits the angle $\omega$ librates around $\pi/2$,
which means the apoapsis always lies above (or below) the $x-y$ plane;
the orbit resembles a conical saucer with inner hole \citep[e.g.][Figure 7]{SambhusSridhar2000}. 
For tubes, conversely, $\omega$ steadily decreases. 
Saucer orbits only exist in the oblate, not prolate case; the condition that the expression 
under the radical in (\ref{eq_R12}) is nonnegative requires that $\calR_z \le \calRsep$,
hence the label ``separatrix''. 

The solution to equation (\ref{eq_dcalRdtau}) can be expressed exactly in terms of the
elliptic cosine \citep[e.g.][Chapter 16; their parameter $m=k^2$, where $k$ is the elliptic 
modulus used as the second parameter in our notation]{AbramovitzStegun}:
\begin{eqnarray}  \label{eq_calR_of_t}
\calR &=& \calRmin + (\calRmax-\calRmin) \times  \\
  &\times& \mathrm{cn}^2 \left(\frac{\CoefK\sqrt{\calRmax-\calRlow}}{2}\,\tau, 
  \sqrt{\frac{\calRmax-\calRmin}{\calRmax-\calRlow}} \right) . \nonumber
\end{eqnarray}
We call the period of full oscillation in $\calR$ the ``precession time.''
It is given by the complete elliptic integral:
\begin{equation}   \label{eq_Tosc}
\tosc = \frac{2}{\pi} \frac{\tprec}{\CoefK\sqrt{\calRmax-\calRlow}} \,
K\left(\sqrt{\frac{\calRmax-\calRmin}{\calRmax-\calRlow}} \right) .
\end{equation}
For orbits not too close to the separatrix, $K \approx \pi/2$.

It is also useful to write down expressions relating $\calRmin$ and $\calRmax$.
For tube orbits,
\begin{equation}
\calRmin=\calRmax - \frac{\calRsep}{1-\calRsep} 
  \frac{(1-\calRmax)(\calRmax-\calR_z)}{\calRmax}.
\end{equation}
It is clear that if $\epsilonp \ll 1$ and $\calRsep \ll \calR \lesssim 1$, 
these two values are quite close to each other, justifying the practice of
approximating the third integral as $L^2$.

For saucer orbits, the relation is simpler:
\begin{equation}  \label{eq_Rminmax_saucer}
\calRmin \calRmax = \calRsep \calR_z.
\end{equation}
In particular, the condition $\calRmin=\calRmax$ gives the fixed-point orbit, 
for which $\omega=\pi/2$ always. 
Figure~\ref{fig_Rminmax} shows the properties of a series of orbits 
with the same $\calR_z < \calRsep$
which start with $\omega=\pi/2$ and with different initial angular momenta 
$\calR_\mathrm{init}$ (left panel), 
and the time evolution of $\calR$ for several saucer orbits with different 
values of $\calH$ (right panel).

\begin{figure}
$$\includegraphics[angle=-90]{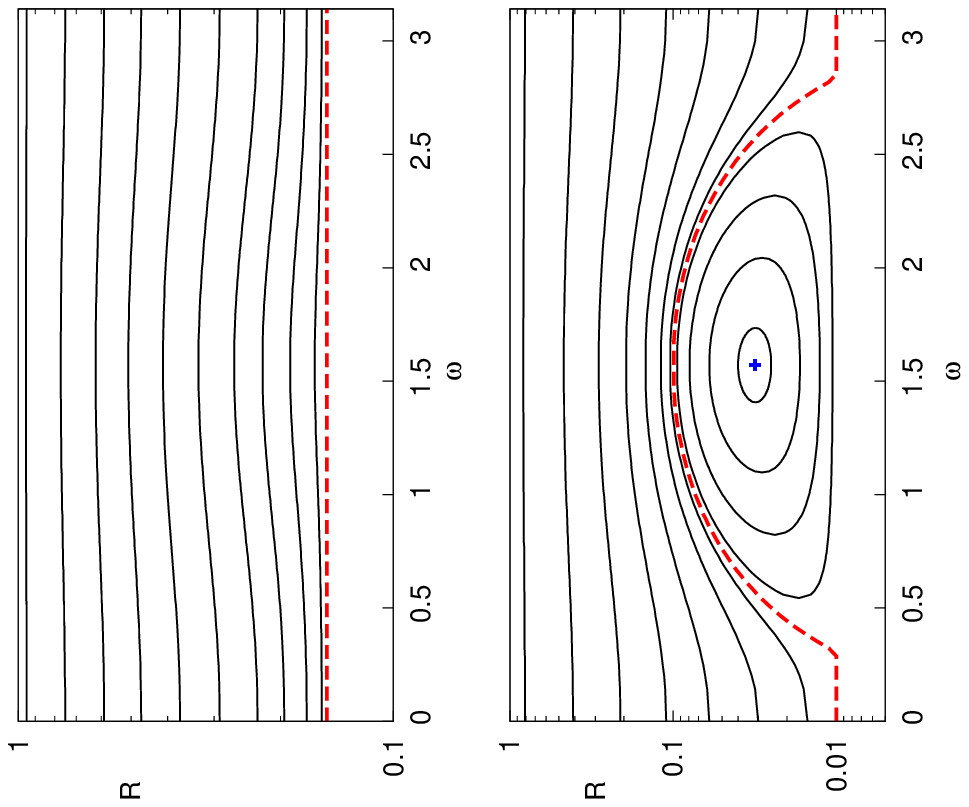}$$
\caption{
$\calR(w)$ (equation \ref{eq_R_of_omega}) for sets of orbits with the same $\calR_z$ and 
different $\calH$. 
Top panel: $\calRsep=0.1,\calR_z=0.15$, which has only tube orbits; 
bottom panel: $\calRsep=0.1, \calR_z=0.01$, which has saucer orbits appearing as cycles
in the lower part of the plot.
The fixed-point orbit is marked by the cross.
In the upper panel, the dashed (red) line is at $\calR=\calR_z$ and in the lower panel
it marks the separatrix.
}\label{fig_poincare}
\end{figure}

Equation (\ref{eq_calH}) can be inverted to express $\calR$ as a function of $\omega$ 
(Figure~\ref{fig_poincare}), which is more convenient for the averaging procedure in the next section:
\begin{subequations}\label{eq_R_of_omega}
\begin{eqnarray}  \label{eq_R_of_omega_a}
\calR(\omega) &=& \frac{\calR_a \pm \sqrt{\calR_a^2-4\calR_z\calR_b(1+\calR_b)}}{2(1+\calR_b)} \;, \\
\calR_b &\equiv& \frac{\calRsep\sin^2\omega}{1-\calRsep} \;,\;\;
\calR_a \equiv \calH+\calR_z + \calR_b(1+\calR_z). \label{eq_R_of_omega_b} \nonumber\\
\end{eqnarray}
\end{subequations}
For tubes only the upper root has physical meaning, while for saucers both roots are valid 
as long as $\sin^2\omega$ is greater than the following threshold:
\begin{equation}  \label{eq_omega_min}
\begin{array}{l}
\displaystyle \sin^2\omega_\mathrm{min} \equiv \frac{1-\calRsep}{\calRsep} \;\times \\
\displaystyle 
  \frac{\calR_z(1-\calH-\calR_z)-\calH + 2\sqrt{-\calH\calR_z(1-\calH-\calR_z)}}{(1-\calR_z)^2}.
\end{array}
\end{equation}
This condition arises from nonnegativity of the expression under the radical in 
(\ref{eq_R_of_omega_a}). Therefore, in a time $\tosc$, $\omega$ varies by $\pi$ for tube orbits 
and by an amount $\le\pi$ for saucers.

\begin{figure}
$$\includegraphics[angle=-90]{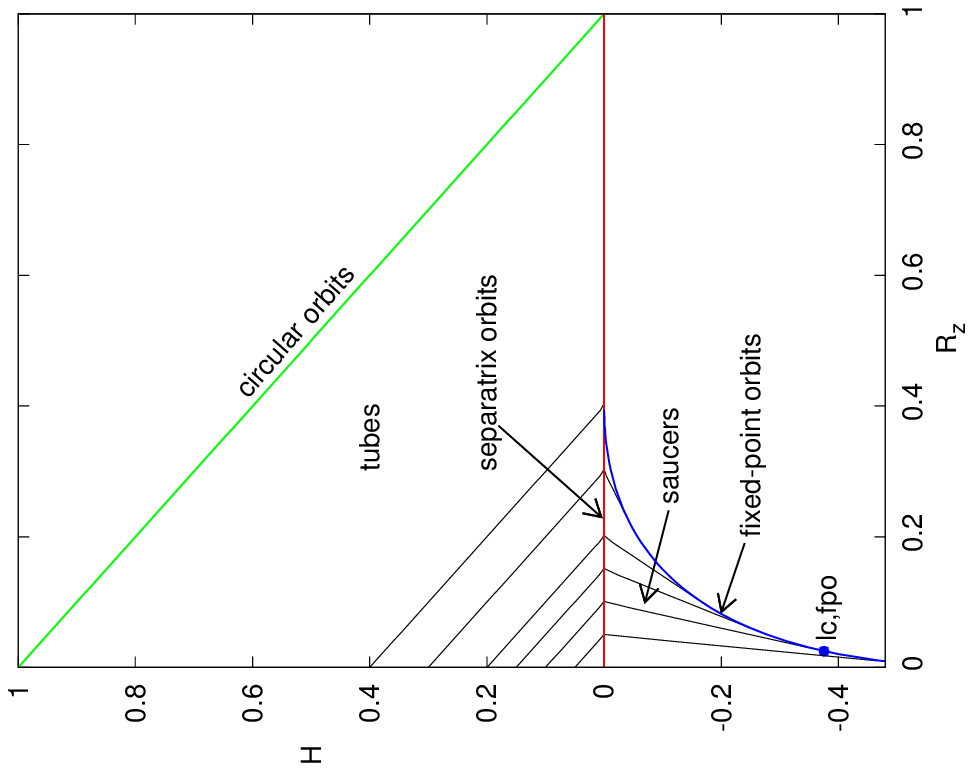}$$
\caption{
Phase space in ($\calR_z, \calH$) coordinates, for $\calRsep=0.4$. 
The region $0<\calH<1-\calR_z$ is occupied by tube orbits, 
and the bottom left corner by saucer orbits; 
the lower boundary (blue) corresponds to fixed-point saucer orbits 
(equation~\ref{eq_FPO}). 
Shown in black are lines of constant minimum angular momentum 
(equation~\ref{eq_capture_boundary}) 
for $\calRcapt=(0.05, 0.1, 0.15, 0.2, 0.3, 0.4)$.
A dot on the intersection of the fixed-point orbit curve and the line of $\calRcapt=0.1$ 
has coordinates given by (\ref{eq_Hlcfpo}).
} \label{fig_phaseplane}
\end{figure}

Finally, we outline the complete phase space in ($\calH, \calR_z$) coordinates 
(Figure~\ref{fig_phaseplane}). 
The boundary between tubes and saucers is $\calH=0$, and 
the other important boundaries are
\begin{subequations}
\begin{equation}  \label{eq_circular_orbit}
\calH=1-\calR_z ,
\end{equation} 
the location of circular orbits ($\calR=1$), and
\begin{equation}  \label{eq_FPO}
\calH=-\frac{\calRsep}{1-\calRsep} \left(1-\sqrt{\frac{\calR_z}{\calRsep}}\right)^2
\end{equation}
is the line of fixed-point saucer orbits, for which $\omega_\mathrm{min}=\pi/2$.
\end{subequations}

A star is captured if it passes near periapsis while having $\calR<\calRcapt$, 
where $\calRcapt\equiv L_\mathrm{lc}^2/I^2 \approx 2r_\mathrm{lc}/a$ is the absorption boundary 
($r_\mathrm{lc}$ is the distance to \sbh\ at which a star is either tidally disrupted or 
captured directly, and $a$ is the semimajor axis). 
In the axisymmetric system, this condition corresponds to $\calRmin<\calRcapt$, 
although not every star satisfying this condition is immediately lost (see 
\S\ref{sec_boundary_cond}).
The lines of constant $\calRmin=\calRcapt$, which mark the \sbh\ capture 
boundary, are straight lines satisfying
\begin{equation}  \label{eq_capture_boundary}
\calH = \left\{\begin{array}{ll}
\calRcapt-\calR_z & \mbox{for tubes} \\
-\frac{(\calRcapt-\calR_z)(\calRsep-\calRcapt)}{(1-\calRsep)\calRcapt} & \mbox{for saucers.} 
\end{array}\right.
\end{equation}
This boundary in the saucer region touches the fixed-point orbit curve (\ref{eq_FPO}) at 
\begin{equation}  \label{eq_Hlcfpo}
\calH_\mathrm{lc,FPO} = -\frac{(\calRsep-\calRcapt)^2}{\calRsep(1-\calRsep)} \;,\quad
\calR_{z,\mathrm{lc,FPO}} = \frac{\calRcapt^2}{\calRsep} 
\end{equation}

In what follows, we will often assume that $\calRsep \ll 1$ 
(or at least not too large), equivalent to assuming that the nuclear flattening is modest.
An isotropic distribution of stars in velocity space corresponds to a distribution 
function which doesn't depend on $\{L,L_z,\omega,\Omega\}$; in these canonical action-angle variables, 
the phase-space volume element is constant and we can compute the proportion of phase space 
occupied by saucer orbits by sampling initial conditions from a uniform distribution in these 
variables and recording the fraction of cases for which $\calH<0$. 
For small $\calRsep$, the fraction of saucer orbits turns out to be approximately $0.4\calRsep$, 
i.e. proportional to the degree of flattening measured by $\epsilonp$ 
(\ref{eq_epsilonp}, \ref{eq_CalHb}).

\section{Fokker-Planck equation}   \label{sec_FP_definition}

We begin by outlining the  general method for deriving the covariant 
Fokker-Planck equation in arbitrary coordinates
\citep{RosenbluthMJ1957}.

The local (position-dependent) Fokker-Planck equation can be
expressed in terms of generalized position- and velocity-space coordinates  
$(x^\alpha, J^\alpha)$, $\alpha = \{1,2,3\}$ as
\begin{eqnarray}  \label{eq_FPlocal}
\frac{\d [\calG f(x^\alpha, J^\alpha, t)]}{\d t} &=& 
- \frac{\d}{\d J^\alpha} (\langle \Delta J^\alpha\rangle \calG f) \;+ \\
  &+& \frac{1}{2} \frac{\d^2}{\d J^\alpha \d J^\beta} 
  (\langle \Delta J^\alpha\Delta J^\beta \rangle \calG f)  \nonumber
\end{eqnarray}
\cite[][equation (5.121)]{DEGN}.
The position and velocity coordinates need not be related to each other in 
any particular way (e.g. they need not be canonically conjugate): 
for instance, one can adopt the integrals of motion as the velocity-space coordinates.
In equation (\ref{eq_FPlocal}), coefficients in angled brackets represent average and 
mean square changes of the corresponding velocity variables per unit time.
Equation~(\ref{eq_FPlocal}) is valid under the assumptions of 
(i) local encounters that change only the velocity but not the position of a star; 
(ii) weak interactions, which allows the collision term to be expanded in powers of 
$\Delta J^\alpha$ up to second order.%
\footnote{\citet{BarorKA2013} argue that retaining only the first two terms in the expansion 
may underestimate the probability of large changes in orbital parameters, especially on  
time scales short compared with the relaxation time.}
Later in this section we will use the orbit-averaged form of this equation, 
which additionally requires that the significant changes in $J^\alpha$ occur on a time
scale  much longer than the orbital period.

The quantity $\calG\equiv a^{1/2}$ is the density of states, 
with $a$ the determinant of the velocity-space 
metric tensor $a_{\alpha\beta}$, so that the squared distance between two points 
whose coordinates differ by $\Delta \boldsymbol{J}$ is given by 
$ds^2 = a_{\alpha\beta}\Delta J^\alpha \Delta J^\beta$, 
and the number of stars in the phase-space volume $d^3J d^3x$ is $dN = \calG f d^3J d^3x$
(generalization to an arbitrary, non-trivial metric in coordinate space is straightforward).
Under a change of coordinates $J^\alpha \to \tilde J^\mu$, 
the coefficients in equation~(\ref{eq_FPlocal}) transform as
\begin{subequations} \label{eq_transform_angled_coefs}
\begin{eqnarray}  \label{eq_transform_angled_coefsa}
\langle \Delta \tilde J^\mu \rangle &=& \langle \Delta J^\alpha \rangle 
  \frac{\d \tilde J^\mu}{\d J^\alpha} +
  \frac{1}{2} \langle \Delta J^\alpha \Delta J^\beta \rangle 
  \frac{\d^2 \tilde J^\mu}{\d J^\alpha \d J^\beta},   \nonumber \\ \\
\label{eq_transform_angled_coefsb}
\langle \Delta \tilde J^\mu \Delta \tilde J^\nu \rangle &=& 
  \langle \Delta J^\alpha \Delta J^\beta \rangle 
  \frac{\d \tilde J^\mu}{\d J^\alpha} \frac{d \tilde J^\nu}{\d J^\beta} 
\end{eqnarray}
\end{subequations}
\cite[][equation (5.120)]{DEGN} and
\begin{equation}
\tilde \calG = \calG \,\det \left\|\frac{\d J^\alpha}{\d\tilde J^\mu}\right\|   .
  \label{eq_transform_jacobian}
\end{equation}

To compute the diffusion coefficients $\langle \Delta J^\alpha \rangle$, 
$\langle \Delta J^\alpha \Delta J^\beta \rangle$, one needs to know the 
distribution function describing the field stars. 
Here we assume that the test and field stars are drawn from the same $f$,
but as is often done, we replace the exact $f$ by an approximation when computing
the diffusion coefficients.
Namely, we neglect changes in the diffusion coefficients caused by 
the evolution, and compute them assuming a field star distribution of the
form $f(\calE)$, where $\calE\equiv -E$ is the binding energy per unit mass. 
In other words, we neglect the anisotropy of the field star distribution.
Consistency with the (spherically-symmetric part of the) mass model 
(\ref{eq_rho_cusp}) requires that
\begin{equation}  \label{eq_distr_function}
f(\calE) = f_0 \calE^{\gamma-3/2}\,,\;\;
f_0 = \frac{\rho_0}{m_\star} \left(\frac{G\Mbh}{r_0}\right)^{-\gamma} \!\!\!\!
  \frac{\Gamma(\gamma+1)}{ (2\pi)^{3/2}\, \Gamma(\gamma-\frac 1 2)} ,
\end{equation}
The assumption of isotropy in the field-star distribution is obviously 
inconsistent with the density model being flattened; 
however, for a small degree of flattening it is a reasonable approximation. 
The flattening may be due to streaming motions, to an elongated velocity ellipsoid
or to both; 
deviations from isotropy due to nuclear flattening are of order   $\epsilonp \ll 1$.

For an isotropic field-star population, the diffusion coefficients  
expressed in terms of $\{v_\|, v_\bot\}$ are well known 
\citep[e.g.][equations~5.23, 5.55]{DEGN}  
and may be transformed to any other 
coordinates using (\ref{eq_transform_angled_coefs}). 
We first adopt the velocity-space variables $J^\alpha = \{ \calE, \calR, \calR_z\}$,
then later replace $\calR$ by $\calH$ which is a true integral of the motion.
The expressions for the local diffusion coefficients in these coordinates are given 
in Appendix \ref{sec_appendix_difcoefs}, equations~(\ref{eq_dif_coefs_local}).

The orbit-averaged Fokker-Planck equation is obtained by 
(i) selecting integrals of motion as the velocity-space coordinates;
(ii) integrating the local Fokker-Planck equation over the phase-space
volume filled by an orbit, assuming that $f(x^\alpha, J^\alpha)$ is a constant 
in this region (Jeans's theorem);
(iii) using the Leibnitz-Reynolds transport theorem to exchange the
order of integration and differentiation.
The result is
\begin{eqnarray}  \label{eq_FPorbitavg}
\frac{\d [\calGav f(J^\alpha,t)]}{\d t} &=& 
- \frac{\d}{\d J^\alpha} (\langle \Delta J^\alpha\rangle_\mathrm{av} \calGav f) \;+ \\
  &+& \frac{1}{2} \frac{\d^2}{\d J^\alpha \d J^\beta} 
  (\langle \Delta J^\alpha\Delta J^\beta \rangle_\mathrm{av} \calGav f)  \nonumber
\end{eqnarray}
\cite[][equation (5.153)]{DEGN},
which has the same form as the local equation (\ref{eq_FPlocal}), but now $f$ is understood to be 
a function of the $J^\alpha$ and of $t$ only, and the diffusion coefficients are averaged 
according to
\begin{eqnarray}  \label{eq_coef_avg}
\langle X\rangle_\mathrm{av}\calGav &\equiv& 
\int_\mathrm{orbit} \langle X\rangle \,\calG\;d^3x .
\end{eqnarray}
Setting $X=1$ in this expression gives the ``density of states'' $\calGav$, 
which relates $f$ to the number of stars in a velocity space volume element
$d^3J$: $dN = f\,\calGav d^3J$.

In the spherically-symmetric case, orbit averaging reduces to a 
one-dimensional integration with respect to radius, $\int_{r_-}^{r_+} X dr/v_r$, 
where $r_-$ and $r_+$ are peri- and apoapsis radii for a given $E$ and $L$;
in other words, averaging reduces to weighting $X$ in proportion to the time 
an orbit spends near $r$. 
In the axisymmetric case, what is traditionally done \citep{Goodman1983, 
EinselSpurzem1999} is to assume that the orbit fills the configuration-space 
region defined by the condition $E - \Phi(R,z) - L_z^2/(2R^2) \ge 0$ where 
($R,z$) are cylindrical coordinates;
in other words, existence of a third integral is ignored.
Then the average turns out to be proportional
simply to $\int\!\!\int X\,dR\,dz$, the integral being taken over this region.

In our case, the distinction between saucer and tube orbits is critical for the
loss-cone problem and we do not want to mix orbits having different values of the third integral.
We must perform the averaging taking into account the shapes of the orbits 
as described in the previous section.
This is most easily done by adopting Delaunay angular variables ($w, \omega, \Omega$) 
as configuration-space coordinates $x^\alpha$. 
If the corresponding actions ($I, L, L_z$) 
were taken as velocity-space coordinates $J^\alpha$, the Jacobian $\calG$ 
of this coordinate system would be unity; 
but since we are using a different set of $J^\alpha$, 
$\calG$ is determined by equation (\ref{eq_transform_jacobian}) for the coordinates 
of choice. 

We split this transformation, and the averaging procedure, into two steps.
Initially we adopt ($\calE, \calR, \calR_z$) as generalized velocities $J^\alpha$ 
and carry out the averaging over the radial phase angle (mean anomaly) $w$, 
obtaining the coefficients for the spherical problem. 
Since these do not depend on $\omega$ or $\Omega$,
averaging over these two angles means simply multiplying by $(2\pi)^2$:
\begin{eqnarray}  \label{eq_coef_avg_spher}
\langle X\rangle_\mathrm{sph}\calGsph \equiv  
  4\pi^2 \int_0^{2\pi} dw\, \langle X\rangle\,
  \det \left\|\frac{\d\{I,L,L_z\}}{\d\{\calE,\calR,\calR_z\}}\right\| .\nonumber \\
\end{eqnarray}

\noindent
Here all coefficients, including the Jacobian (explicitly written as a determinant), 
are understood to be functions of ($\calE, \calR, \calR_z$). 
In the case of a Keplerian potential $\Phi(r)=-G\Mbh/r$ (i.e. neglecting the 
contribution of the stars), and assuming a power-law density profile for the stars 
with index $\gamma=3/2$, the field star distribution function (in our isotropic
approximation)  is a constant $f_0$ (equation \ref{eq_distr_function}), 
and the diffusion coefficients can be expressed in terms of elementary functions as
\begin{subequations}  \label{eq_difcoefsRRz}
\begin{eqnarray} 
\langle\Delta\calE\rangle_\mathrm{sph}  &=&
  \CoefA\,\calE\,\left(\frac{2}{3}\frac{m}{m_\star}-1\right) , \\
\langle\Delta\calR\rangle_\mathrm{sph}  &=&
  \CoefA\,\frac{29-66\calR}{15} ,\\
\langle\Delta\calR_z\rangle_\mathrm{sph}  &=&
  \langle\Delta\calR\rangle_\mathrm{sph}\, \frac{\calR_z}{\calR} 
  + \frac{\CoefA}{30\calR} \big\{9\calR(\calR-3\calR_z) \nonumber  \\
  &+& 29(1-\calR)[(\calR-\calR_z)\cos^2\omega-\calR_z]\big\}\qquad \phantom{.} \\
\langle(\Delta\calE)^2\rangle_\mathrm{sph} &=&
  \CoefA\,\calE^2\;\frac{32-4\sqrt{\calR}}{15\sqrt{\calR}} , \\
\langle(\Delta\calR)^2\rangle_\mathrm{sph} &=&
  \CoefA\,\calR\,\frac{58+32\sqrt{\calR}-90\calR}{15} , \\
\langle(\Delta\calR_z)^2\rangle_\mathrm{sph}&=&
  \langle(\Delta\calR)^2\rangle_\mathrm{sph} \left(\frac{\calR_z}{\calR}\right)^2 
  + \frac{2\CoefA}{15} (\calR-\calR_z) \nonumber \\
  &\times& \frac{\calR_z}{\calR} \big[ 9\calR + 29(1-\calR) \cos^2\omega\big] , 
  \label{eq_difcoefRz2} 
\end{eqnarray} 
\begin{eqnarray} 
\langle\Delta\calE \Delta\calR \rangle_\mathrm{sph}  &=&
  \CoefA\,\calE\,\frac{32\sqrt{\calR}(1-\sqrt{\calR})}{15} , \\
\langle\Delta\calE \Delta\calR_z\rangle_\mathrm{sph} &=&
  \langle\Delta\calE \Delta\calR \rangle_\mathrm{sph}\frac{\calR_z}{\calR} , \\
\langle \Delta\calR \Delta\calR_z\rangle_\mathrm{sph} &=&
  \langle \Delta \calR^2 \rangle_\mathrm{sph}\frac{\calR_z}{\calR}
\end{eqnarray}
\end{subequations}
where
\begin{equation}  \label{eq_difcoefA}
\CoefA \equiv 16\pi^2G^2m_\star^2\ln\Lambda\,f_0 .
\end{equation}
Here $m_\star$ is the mass of a field star (scatterer) and $m$ is the mass 
of a test star (whose evolution is described by the Fokker-Planck equation). 
We set $m_\star=m$ in what follows. 

The density of states $\calGsph$ is split into two factors:
\begin{equation}  \label{eq_densityofstates}
\calGsph = \frac{\calG_\calE}{2\sqrt{\calR\calR_z}} \;,\quad 
  \calG_\calE \equiv \frac{\sqrt{2}\,\pi^3\, (G\Mbh)^3}{\calE^{5/2}}
\end{equation}
 so that 
$\int_0^1 d\calR \int_0^\calR d\calR_z\,\calGsph = \calG_\calE$ 
is the phase space volume element that transforms $f$ to the number density 
of stars $\calN$ in the spherical geometry:
\begin{equation}  \label{eq_NofER}
\calN(\calE,\calR)\, d\calE d\calR = \calG_\calE(\calE)\, f(\calE,\calR)\, d\calE d\calR.
\end{equation}

These coefficients are based on the usual approximation of uncorrelated two-body encounters. 
The effect of resonant relaxation \citep{RauchTremaine1996} would be to enhance the diffusion 
coefficients for $\calR$ and $\calR_z$; we defer a discussion of this until \S\ref{sec_Nbody}.

Finally, the diffusion coefficients are expressed in terms of 
$\{\calE, \calH, \calR_z\}$ by substituting $\calH(\calR,\calR_z,\omega)$ from 
equation (\ref{eq_calH}), transforming the above coefficients under this substitution 
according to equation (\ref{eq_transform_angled_coefs}), and averaging over the 
argument of periastron (i.e. the precession phase) $\omega$:
\begin{equation}  \label{eq_coef_avg_omega}
\langle X\rangle_\mathrm{av}\calGav = \frac 1 \pi 
  \int_{\omega_\mathrm{min}}^{\pi-\omega_\mathrm{min}} d\omega\, \langle \tilde X\rangle_\mathrm{sph}\, 
  \frac{\d\calR}{\d\calH} \, \calGsph .
\end{equation}
We have explicitly written the Jacobian of transformation (\ref{eq_transform_jacobian}), 
and the coefficients with tildes are transformed from $\langle X \rangle_\mathrm{sph}$ using 
(\ref{eq_transform_angled_coefs}). This is possible because the last 
transformation does not depend on $w$ or $\Omega$. 
The limits of integration in (\ref{eq_coef_avg_omega}) are 
between 0 and $\pi$ for tubes and between $\omega_\mathrm{min}$ and 
$\pi-\omega_\mathrm{min}$ for saucers (equation~\ref{eq_omega_min}).
From here on, we omit the subscript for the averaged coefficients.

The averaging must be done numerically, however, we can derive asymptotic 
expressions for the  diffusion coefficients in the large- and small-$\calR$ regimes. 
In the former case, when $\calR\gg\calRsep$, 
$\calR\approx\const = \calH+\calR_z$, so we trivially get 
$\langle\Delta\calH^2\rangle = \langle(\Delta\calR)^2\rangle - 
2\langle\Delta\calR\Delta\calR_z\rangle + \langle(\Delta\calR_z)^2\rangle$, 
$\langle\Delta\calH\Delta\calR_z\rangle = \langle(\Delta\calR)^2\rangle - 
\langle\Delta\calR\Delta\calR_z\rangle$. 

In the limit of small $\calR$, i.e. close to the capture boundary, 
the asymptotic behavior of the coefficients is different in the tube and saucer regions. 
In the case of tube orbits with $\calH+\calR_z \ll \calRsep$, the asymptotic expressions are
\begin{equation}  \label{eq_coef_asympt_SAT}
\langle\Delta\calH^2\rangle \approx \CoefA \frac{58}{15}\frac{\calRsep}{\CoefLn} \;,\quad
\calGav \approx \frac{2\CoefLn}{\pi} \frac{\calG_\calE}{2\sqrt{\calRsep\calR_z}} \;,
\end{equation}
where $\CoefLn \equiv 2\ln2 - \frac{1}{2}\ln(\calR_z/\calRsep)$; 
the other two diffusion coefficients are smaller that $\langle\Delta\calH^2\rangle$ 
by a factor $\sim \calR_z/\calRsep$.
For saucer orbits with $\calR_z \ll \calRsep$ and not too close to either the separatrix 
or the fixed-point orbit, the asymptotic expressions for the diffusion coefficients are 
estimated to within $\mathcal{O}(1)$ as
\begin{equation}
\langle(\Delta\calR_z)^2\rangle \approx 2\CoefA \left(1+\frac{\calH}{\calRsep}\right)\,\calR_z \;,\quad
\langle\Delta\calH\Delta\calR_z\rangle \approx 4\CoefA \calR_z \;,  \nonumber
\end{equation}
\begin{equation}  \label{eq_coef_asympt_SAU}
\langle\Delta\calH^2\rangle \approx \langle\Delta\calH^2\rangle \,\frac{\calRsep}{\calR_z} \;,\quad
\calGav \approx \frac{\calG_\calE}{\sqrt{2\calRsep\calR_z}}. \\
\end{equation}

Finally, it is convenient to cast the orbit-averaged
Fokker-Planck equation (\ref{eq_FPorbitavg}) into flux-conservative form:
\begin{equation}  \label{eq_FPflux}
\frac{\d (\calGav f)}{\d t} = -\frac{\d F^\alpha}{\d J^\alpha} \;,\quad
  F^\alpha \equiv -\Adv^\alpha f - \Dif^{\alpha\beta} \frac{\d f}{\d J^\beta} .
\end{equation}
The drift and diffusion coefficients are 
\begin{subequations}  \label{eq_difcoefs_definition}
\begin{eqnarray}  \label{eq_difcoefs_definition_a}
\Adv^\alpha &=& -\calGav \langle \Delta J^\alpha \rangle 
  + \frac{\d}{\d J^\beta} \left(\frac{1}{2} \calGav 
  \langle \Delta J^\alpha \Delta J^\beta \rangle \right) , \quad\phantom{,} \\
\Dif^{\alpha\beta} &=& \frac{1}{2} \calGav 
  \langle \Delta J^\alpha \Delta J^\beta \rangle . \label{eq_difcoefs_definition_b}
\end{eqnarray}
\end{subequations}
This form is convenient from both from a computational and a physical point of view, 
since
(i) it manifestly conserves the total number of stars, and
(ii) drift coefficients $\Adv_\calR$ and $\Adv_{\calR_z}$ must be zero, 
which may be demonstrated explicitly, or inferred from the natural requirement that 
(in the absence of capture) collisions tend to isotropize $f$, i.e. in the steady state $f$
should not depend on $\calR$ or $\calR_z$.

\section{Boundary conditions}   \label{sec_boundary_cond}

As is well known, the steady-state loss rate in the spherical geometry can only be
inferred from the orbit-averaged Fokker-Planck equation if a certain condition is satisfied:
the change in angular momentum over one radial period due to encounters must
be small for orbits near the loss cone.
Far enough from the \sbh, this condition is violated (the ``pinhole'' or ``full-loss-cone''
regime) and the orbit-averaged approximation is not valid.
This problem is dealt with by returning to the local $(r,v)$ Fokker-Planck equation 
and allowing the phase-space density to vary along orbits, from zero at the edge
of the capture sphere to some finite value at apoapsis \citep{CohnKulsrud1978}.
The result is a boundary condition for the Fokker-Planck equation
that specifies the value and slope of $f(\calR)$ at
$\calR=\calRcapt$ in terms of its average value at the (fixed) energy $\calE$.

No such exact analysis has ever been carried out in nonspherical (axisymmetric, triaxial)
geometries, nor will we do so here.
Instead we will be satisfied with a more heuristic derivation of the relation between
$f$ and $\calF$ at the loss cone boundary.
Our analysis will be similar in spirit to that of \citet{MagorrianTremaine1999},
but, as we will see, with some important differences.

\subsection{The spherical problem}  \label{sec_boundary_cond_spher}

We begin by reviewing the boundary conditions in the spherical geometry.
The evolution equation in terms of $\calE$ and $\calR$ can be derived by
integrating equation (\ref{eq_FPflux}) over $\calR_z$ from 0 to $\calR$. 
We introduce two, two-dimensional quantities: the number density of stars,
\begin{eqnarray}
\calN(\calE,\calR)\, d\calE d\calR &\equiv& 
  \int_0^{\calR}\!\! d\calR_z\, \calGsph(\calE,\calR,\calR_z)\, 
  f(\calE,\calR,\calR_z)\, d\calE d\calR  \nonumber  \\
&=& 4\pi^2 \trad(\calE)\Lcirc^2(\calE)\,f(\calE,\calR)\, d\calE d\calR  \nonumber \\
&=& \calG_\calE(\calE)\,f(\calE,\calR)\, d\calE d\calR,  \label{Equation:NofER}
\end{eqnarray}
and the flux per unit energy in the $\calR$-direction:
\begin{eqnarray}
\calF_\calR(\calE,\calR)\, d\calE &=& -\int_0^{\calR} d\calR_z \, \calGsph\, 
\frac{\langle(\Delta\calR)^2\rangle}{2} \, \frac{\d f}{\d \calR}\, d\calE  \nonumber \\
&=& \frac{\langle(\Delta\calR)^2\rangle}{2} \frac{\d\calN(\calE,\calR)}{\d\calR}\, d\calE.
  \label{Equation:FluxRofE}
\end{eqnarray}
The relation between $\calF_\calR$ and the fluxes appearing in equation~(\ref{eq_FPflux}) 
is that the former is the integral, over the entire loss cone boundary, of the component 
of $F^\alpha$ normal to that boundary, in the subspace $\calE=\mathrm{const}$. 
In the spherical case, the capture boundary in the $\calR-\calR_z$ plane is defined by 
$\calR=\calRcapt$, and so $\calF_\calR = \int_0^\calR d\calR_z\,F^{\calR}$. 

As emphasized by \citet{FrankRees1976}, in the spherical geometry, 
changes in angular momentum are expected to dominate the capture rate.
Setting $\calE=$ constant, we consider one-dimensional diffusion in $\calR$,
which obeys
\begin{eqnarray}  
&&\frac{\d \calN(\calR,t)}{\d t} = -\frac{\d\calF_\calR}{\d\calR} \nonumber \\
&&= -\frac{\d}{\d\calR} \left(\calN\langle\Delta\calR\rangle\right)
  +\frac 1 2 \frac{\d^2\calN }{\d\calR^2}
  \left(\calN\langle(\Delta\calR)^2\rangle  \right) \nonumber \\
&&\approx\frac 1 2  \frac{\d}{\d\calR}\left(\langle(\Delta\calR)^2\rangle
  \frac{\d \calN} {\d\calR}\right). \label{eq_dif_1d}
\end{eqnarray}
The final expression uses the result that for small $\calR$, 
$\langle\Delta\calR\rangle = \frac12(\d/\d\calR)\langle(\Delta\calR)^2\rangle$.
Also in this limit, $\langle(\Delta\calR)^2\rangle \propto\calR$, 
and we can write
\begin{equation}  \label{eq_d_definition}
\Dif(\calE) \equiv \lim_{\calR\to 0} \frac{\langle(\Delta\calR)^2\rangle}{2\calR} ,
\end{equation}
an inverse, orbit-averaged relaxation time. 

If we adopt the approximation $\langle(\Delta\calR)^2\rangle/2=\Dif\calR$ 
over the entire range of $\calR$,
then equation (\ref{eq_dif_1d}) is mathemathically equivalent to the heat 
conduction equation in a circular domain \citep{Ozisik1993}.
The steady-state solution is $\calN\propto\ln R+\mathrm{const}$.
The natural boundary condition at $\calR=1$ is $\calF_\calR=0$, 
which in this approximation translates to $\d\calN/\d\calR=0$ 
(in the more exact treatment, $\langle(\Delta\calR)^2\rangle$ itself tends to zero at $\calR=1$).
The boundary at $\calR=\calRcapt$ is responsible for capture, however, 
simply setting $\calN(\calRcapt)=0$ is not always appropriate. 
As discussed by \citet{LightmanShapiro1977}, there are two regimes 
characterizing the behavior of $\calN$ near loss cone boundary, depending on the 
ratio $q$ between the radial period and the time to random-walk out of the loss cone
(that is, to change $\calR$ by order of $\calRcapt$):
\begin{equation}  \label{eq_q_spher}
q(\calE) \equiv \frac{\trad}{T_\mathrm{rw}}\;,\quad
T_\mathrm{rw} \equiv \frac{\calRcapt^2}{\langle(\Delta\calR)^2\rangle/2} = \frac{\calRcapt}{\Dif}.
\end{equation}
The case $q \ll 1$ is the ``empty loss cone'' regime, since a star scattered into 
the region $\calR\le\calRcapt$ is swallowed much faster than encounters
can scatter it back out. 
The opposite situation, $q\gg 1$, is the
``full-loss-cone'' regime, because diffusion is so rapid that even capture of 
stars (near periapsis)  does not substantially diminish the population of loss
cone orbits away from periapsis.

It turns out that much of the flux into the \sbh\ comes from stars at energies where
$q\approx 1$,
so the behavior of the solution at the transition between the two regimes is important.
Let $\calN_\mathrm{lc}\equiv \calN(\calRcapt)$, and define 
$\calF_\mathrm{lc}=-\calF_\calR(\calRcapt)$,
the flux through the boundary 
(i.e. the number of stars captured per unit time per unit energy). 
In a steady state, equation (\ref{eq_dif_1d}) implies that 
$\calF_\calR$ is independent of $\calR$ near $\calRcapt$.
We can express the inner boundary condition in a general way as 
\begin{equation}  \label{eq_boundary_cond_spher}
\calF_\mathrm{lc} = \alpha^{-1}\calN_\mathrm{lc}\Dif
\end{equation}
where all quantities are understood to be functions of $\calE$, or equivalently of
$q(\calE)$.
After usinq equations (\ref{Equation:NofER}) and (\ref{Equation:FluxRofE}) to express $\calN$ 
and $\calF_\calR$ in terms of $f$ and $\d f/\d\calR$, equation (\ref{eq_boundary_cond_spher}) 
is seen to be a boundary condition of the Robin type 
(linear combination of function and its derivative) \citep{Eriksson1996}.
The dimensionless coefficient $\alpha$ can be derived
by returning to the local (non-orbit-averaged) Fokker-Planck equation and 
determining how $f$ varies with radial phase assuming $f=0$ 
at periapsis \citep{Baldwin1972}.
\citet{CohnKulsrud1978} produced a numerical solution in the spherical geometry
and proposed the approximation
$\alpha = 0.186q+0.824\sqrt{q}$ for $q\le 1$ and 
$\alpha= q$ for $q\ge 1$.
An exact solution exists to the same set of equations solved by Cohn \& Kulsrud:
\begin{equation}  \label{Equation:Besselalpha}
\alpha(q) = \frac{q}{\xi(q)}, \ \ \xi(q) = 1 - 4\sum_{m=1}^\infty \frac{e^{-\alpha_m^2q/4}}{\alpha_m^2}
\end{equation}
\citep{DEGN},
where $\alpha_m$ are consecutive zeros of the Bessel function $J_0$.
Equation (\ref{Equation:Besselalpha}) is unwieldy; a good approximation is 
\begin{equation}  \label{eq_alpha_q}
\alpha(q) =(q^4+q^2)^{1/4} 
\end{equation}
which has the asymptotic forms
\begin{equation}  \label{Equation:alphaasympt}
\alpha \rightarrow 
\begin{cases} 
\sqrt{q} + q^{5/2}/4 &\mbox{if } q \ll 1 \\
q + 1/(4q) & \mbox{if } q \gg 1. 
\end{cases}
\end{equation}
These  expressions differ most strongly near $q=1$, where the 
exact solution (\ref{Equation:Besselalpha}) gives $1.195$.
Cohn \& Kulsrud's approximation is $\alpha(1)=1$ while equation (\ref{eq_alpha_q}) gives $1.189$.

In terms of $\alpha$, the variation of $\calN$ with $\calR$ near the loss cone boundary is
\begin{subequations} \label{eq_f_near_lc_spher}
\begin{eqnarray}  \label{eq_f_near_lc_sphera}
\calN(\calR) &\approx& \Dif^{-1}\calF_\mathrm{lc}\,\left(\alpha+\ln\frac{\calR}{\calRcapt}\right)\nonumber \\
 &\approx& \Dif^{-1}\calF_\mathrm{lc}\,\ln\frac{\calR}{\calR_0}, \\
 \calR &\ge& \calR_0\equiv \calRcapt\exp(-\alpha).  \label{eq_R0}
 \label{eq_f_near_lc_spherb}
\end{eqnarray}
\end{subequations}
Here $\calR_0\le \calRcapt$ plays the role of the effective absorbing boundary 
at which $\calN=0$.

It is convenient to introduce the ``draining rate'' of a uniformly-populated loss cone:
\begin{equation}  \label{eq_Fdrain}
\calFdrain \equiv \frac{\calN_\mathrm{lc}\,\calRcapt}{\trad}
= \frac{\calG_\calE f_\mathrm{lc}\,\calRcapt}{\trad}  
= 4\pi^2 L_\mathrm{lc}^2 f_\mathrm{lc}. 
\end{equation}
This is the capture rate that results if the following two conditions are satisfied:
(i) the distribution function is constant inside the loss cone, with value $f_\mathrm{lc}$;
(ii) the volume of the loss cone (per unit energy), $\calG_\calE\calRcapt$, 
is emptied every radial period. 
As is apparent from its definition, 
$\calFdrain$ does not depend on the diffusion coefficient $\Dif$.

We can rewrite the boundary condition (\ref{eq_boundary_cond_spher}) in 
terms of $\calFdrain$ as
\begin{equation}  \label{eq_Flux_sph}
\calF_\mathrm{lc} = \frac{q}{\alpha}\,\calFdrain .
\end{equation}
In the full-loss-cone regime, $q\approx \alpha$ and the capture rate is 
$\calF_\mathrm{lc} \approx \calFdrain$; otherwise $q/\alpha<1$, reflecting the fact that the phase
space density decreases to zero at some $\calR=\calR_0\lesssim\calRcapt$ (\ref{eq_R0}).

Define $\overline{\calN}(\calE)$ to be the integral of $\calN(\calE,\calR)$ over angular momenta:
\begin{equation}  \label{eq_Navg}
\overline \calN(\calE) \equiv \int_{\calRcapt}^1 \calN(\calE,\calR)\,d\calR.
\end{equation}
Roughly speaking, $\overline{\calN}$ is the quantity that would be inferred from an observed
radial density profile, for instance, via Eddington's formula \citep[e.g.][equation~3.47]{DEGN}.
If we extrapolate the logarithmic solution (\ref{eq_f_near_lc_spher}) to all $\calR$, 
we can relate $\calN(\calR)$ and the capture rate $\calF_\mathrm{lc}$ to
$\overline \calN$:
\begin{subequations}
\begin{eqnarray}    \label{eq_f_global_spher}
\calN(\calR) &\approx& 
  \frac{\alpha + \ln(\calR/\calRcapt)}{\alpha + \ln(1/\calRcapt)-1} \,\overline\calN,  \\
\calF_\mathrm{lc} &\approx& 
  \frac{\Dif\ \overline \calN}{\alpha + \ln(1/\calRcapt)-1}.  \label{eq_capt_rate_spher} 
\end{eqnarray}
\end{subequations}
In the full-loss-cone regime, $q\gg 1$, the distribution function is almost isotropic 
($\calN(\calR) \approx \overline \calN$), and
\begin{equation}  \label{eq_FFLC}
\calF_\mathrm{lc} \rightarrow \frac{{\overline\calN} \calRcapt}{\trad}.
\end{equation}
That is: the full volume of the loss cone is consumed every radial period, 
and $\calFdrain$ and $\calF_\mathrm{lc}$ are equivalent. 
This also means that the exact value of the diffusion coefficient or even the 
very process responsible for shuffling orbits in $\calR$ does not affect the 
capture rate, as long as it is efficient enough to keep the loss cone full.
In the opposite limit of $q\ll 1$,
\begin{equation}  \label{eq_FELC}
\calF_\mathrm{lc} \rightarrow \frac{q}{\ln(1/\calRcapt)} \frac{{\overline\calN} \calRcapt}{\trad}.
\end{equation}
Now the capture rate is limited by diffusion from larger 
$\calR$, that is, by the gradient of $\calN$ near $\calRcapt$. 
The distribution function is  depleted at small $\calR$.

Here we note a distinction that will be important when discussing the axisymmetric problem.
Equation (\ref{eq_Fdrain}) expresses $\calFdrain$ in terms of the value of $\calN$ 
at the loss cone boundary.
In the spherical case, the full-loss-cone boundary condition ($\alpha\approx q\gg1$) 
implies $\calN_\mathrm{lc} \approx \overline \calN$ (equation \ref{eq_f_global_spher}).
This is because the same process -- gravitational scattering -- 
is responsible both for populating loss-cone orbits uniformly with respect 
to radial phase and for driving the global shape of $f(\calR)$ towards isotropy.
As we will see, the same is not necessarily true in the axisymmetric geometry,
because these two actions are driven by different processes: 
the latter is always attributed to two-body relaxation (scattering), 
but the former may also be driven by regular precession.
In what follows, we use the terms ``empty'' and ``full'' loss-cone regimes to distinguish 
between the cases when the flux $\calF_\mathrm{lc} \ll \calFdrain$ and $\approx \calFdrain$,
respectively, whatever the global shape of the solution.

In the spherical problem, the transition between the two regimes is naturally defined as 
$\alpha \approx q = \ln 1/\calRcapt$ (so that expressions (\ref{eq_FFLC}) and 
(\ref{eq_FELC}) are equal). 
Sometimes another definition is used, based on the requirement that the draining rate 
equals the repopulation rate from nearby regions in phase space, implying $q=1$.
In \S\ref{sec_Nbody} we denote the energy of the former transition as 
$\calE_\mathrm{global}$ and the latter as $\calE_\mathrm{local}$, 
in reference to the fact that these conditions are based either on the global 
shape of the solution or  on its local properties near the capture boundary.

Returning to the time-dependent equation (\ref{eq_dif_1d}),
if $\langle(\Delta\calR)^2\rangle=2\Adv\calR$,
an analytic solution exists in terms of Bessel functions
\citep{MilosMerritt2003,MerrittWang2005}.
If we take $\calN(\calR)=\Theta(\calR-\calRcapt)$ as the initial 
condition, where $\Theta$ is the Heaviside step function, then the logarithmic 
profile is established after $\Delta t\approx 10^{-2}\Dif^{-1}$, and
the flux $\calF_\mathrm{lc}$, after the initial transient,
is well described by the quasi-steady-state value (\ref{eq_capt_rate_spher}).
Numerical solutions to equation (\ref{eq_dif_1d})
without the simplifying assumption $\langle(\Delta\calR)^2\rangle/\calR=\const$
are found to match the analytical solution very well (to within a few percent).
In \S~\ref{sec_Nbody} we will refer to both the quasi-stationary flux 
value and the time-dependent solution. 

One should keep in mind that the capture rate in the time-dependent case 
may be substantially higher than the stationary flux for $t\ll 1/\Dif$, at least
in the empty-loss-cone regime; sometimes the ratio can be more than an order 
of magnitude  \cite[e.g.][]{MilosMerritt2003}. 
This, however, depends critically on the details of the initial distribution.
If instead of a step-function at $\calRcapt$ one starts from a profile with a 
larger area $\calR_\mathrm{depl} \gg \calRcapt$ where the distribution function 
has been depleted, for example, as a result of a binary SMBH scattering away stars with 
periapsides smaller than the binary separation, then initially the capture rate is,
conversely, much lower than the stationary flux \citep{MerrittWang2005}.

\subsection{The axisymmetric problem}  \label{sec_boundary_cond_axi}

We first summarize the various time scales in the axisymmetric geometry.
The three times defined above that characterize motion in the smooth potential satisfy
\begin{equation}
\trad \ll \tprec \lesssim \tosc 
\end{equation}
where $\trad$ (equation \ref{eq_Trad}) is the radial (Keplerian) period,
$\tprec$ (equation \ref{eq_Tprec}) is the approximate time for $\omega$ to change
by $2\pi$ due to the spherically-distributed mass,
and $\tosc$ (equation \ref{eq_Tosc})  is the oscillation time for $\calR$ due to torquing
from the axisymmetric component of the potential.
The latter inequality is strongest
for saucer orbits that are near the separatrix and which precess very slowly
(Figure \ref{fig_Rminmax}).

The two-body relaxation time can be estimated from the diffusion 
coefficients by taking the inverse of the common dimensional factor $\CoefA$, equation (\ref{eq_difcoefA}):
\begin{equation}  \label{eq_Trel}
\trel \equiv \CoefA^{-1} = \tprec\,\frac{\Mbh}{m_\star} \frac{4\sqrt{2}}{9\pi\ln\Lambda}.
\end{equation}
This time is smaller by a factor 0.87 than the more standard definition of the relaxation 
time in terms of local density and velocity dispersion (cf.\ \citet{DEGN}, equation~5.61).
Clearly $\trel\gg\tprec$.

\citet{MagorrianTremaine1999} used the term ``loss wedge'' to describe 
the set of orbits which can be captured by the \sbh\ in the absense of relaxation, 
i.e. if their angular momentum falls below $\calRcapt$ at some phase of the
precession cycle, or, equivalently, if $\calR_\mathrm{min}\le\calR_\mathrm{lc}$.
The name reflects the fact that this region is elongated in the $\calH$ direction much more 
than in $\calR_z$ (Figure~\ref{fig_phaseplane}): its boundary is defined by setting 
$\calRmin(\calH,\calR_z)=\calRcapt$ in equation (\ref{eq_capture_boundary}).
In what follows, we define $f_\mathrm{lw}$ as the value of the distribution function at the 
loss wedge boundary, which we approximate to be constant throughout the loss wedge,
and $\calF_\mathrm{lw}$ as the capture rate of stars per unit energy:
\begin{equation}  \label{eq_capt_rate_total_flux}
\calF_\mathrm{lw} \equiv -\int_{\calH_\mathrm{lc,FPO}}^{\calRcapt}\!\! F^{\calR_z}\, d\calH -
  \int_0^{\calRcapt} (F^{\calH}_\mathrm{tube} - F^{\calH}_\mathrm{saucer})\, d\calR_z\;.
\end{equation}
Here $F^{\calH}$ and $F^{\calR_z}$ are fluxes defined in equation (\ref{eq_FPflux}), 
$\calH_\mathrm{lc,FPO}$ is the lowest possible value of $\calH$ (\ref{eq_Hlcfpo}) 
for orbits outside the loss wedge,
and the two terms in the last integral give the contributions to the capture rate from 
the ``downward'' flux in the $\calH$ direction in the tube region of phase plane and 
the ``upward'' flux in the saucer region (see Figure~\ref{fig_phaseplane}). 
Signs are chosen so that $\calF_\mathrm{lw}$ is the positive rate of capture.

We now argue that it is the flux in the $\calR_z$ direction in the saucer region that 
provides the main contribution to the total capture rate, in the case  
$\calRcapt \ll \calRsep \ll 1$.
From equations (\ref{eq_coef_asympt_SAU}), (\ref{eq_difcoefs_definition_b}) we see that 
$\Dif^{\calR_z\calR_z} \sim \Dif^{\calH\calR_z} \sim (\calR_z/\calRsep)\Dif^{\calH\calH}$.
The capture boundary (\ref{eq_capture_boundary}) is almost parallel to the vertical 
($\calH$) axis in this region, with the slope $\calRsep/\calRcapt \gg 1$. 
If we assume that $f(\calR_z,\calH)$ has a certain gradient perpendicular to the 
capture boundary line, then its derivatives are in a similar relation: 
$\d f/\d \calR_z : \d f/\d \calH \sim \calRsep : \calRcapt$.
The fluxes $F^{\calR_z}$ and $F^{\calH}$ in (\ref{eq_FPflux}) are then comparable in 
magnitude; however, in equation (\ref{eq_capt_rate_total_flux}) the former flux 
is integrated in $d\calH$ on an interval of length $\sim \calRsep$, while the latter is 
integrated in $d\calR_z$ on an interval of $\calRcapt$. Therefore, the contribution 
from the flux in the $\calR_z$ direction is the largest in the saucer region.
From similar arguments we estimate that in the tube region the flux in the $\calH$ direction 
is dominant. 
Finally, if $\calRcapt \ll \calRsep$, most of the loss wedge lies in the saucer 
orbit region, so it gives the largest overall contribution to the total capture rate
$\calF_\mathrm{lw}$. (This is true only asymptotically; as shown in the next section, even for 
$\calRcapt/\calRsep \sim 0.03$ the tube and saucer regions give roughly equal contribution).

The number of stars (per unit energy) inside the loss wedge is given by%
\footnote{Note that here $\calN_{<\mathrm{lw}}$ denotes the \textit{integral} of 
the distribution function over the loss region, not its \textit{value} at the boundary 
as in (\ref{Equation:NofER}).}
\begin{equation}  \label{eq_Nlw}
\calN_{<\mathrm{lw}} \equiv \int\!\!\int \calGav f\,d\calH d\calR_z 
  \approx \calG_\calE f_\mathrm{lw}\, \sqrt{\calRsep\calRcapt} \;,
\end{equation}
where the boundaries of the region of integration are given by $\calR_z\le \calRcapt$ 
and equation (\ref{eq_capture_boundary}), and we have taken $f$ to be constant ($f_\mathrm{lw}$)
within this region and used the asymptotic expression (\ref{eq_coef_asympt_SAU}) for $\calGav$ 
in the saucer region (which gives the main contribution to the integral in the case 
$\calRcapt \ll \calRsep$).
The number of stars which are instantaneously inside the loss cone ($\calR<\calRcapt$)
is essentially the same as in the spherical case: $\calG_\calE f_\mathrm{lw}\calRcapt$.
This is an expected result:
in the axisymmetric case, the effective volume of the loss region increases by a factor
$\sim \sqrt{\calRsep/\calRcapt}$, but the probability for any star inside this loss region 
 having an angular momentum less than the capture threshold decreases by the same factor.

We are now in a position to derive the boundary condition, i.e., to relate the capture rate 
$\calF_\mathrm{lw}$  to the value of $f$ at the boundary of the loss wedge, $f_\mathrm{lw}$.
An orbit inside the loss \textit{wedge} is captured if its instantaneous value of $\calR$ 
is less than $\calRcapt$, i.e., if it is in the loss \textit{cone}, during periapsis passage.
Here, as in the spherical case, there are two possible regimes. 
If the radial period is short compared with the time required for an orbit to pass the minimum 
of its precession cycle while having $\calR<\calRcapt$, then every orbit in the loss wedge 
will be captured in a time no longer than one precession period. 
We call this the ``empty loss wedge'' regime.
The rate of consumption of stars per unit energy, $\calF_\mathrm{lw}$, 
is then given by the number of stars inside the loss wedge, $\calN_{<\mathrm{lw}}$, 
divided by their lifetime on these orbits, $\tosc$:
\begin{equation}  \label{eq_flux_axi_empty}
\calF_\mathrm{drain,lw} \approx \frac{\calN_{<\mathrm{lw}}}{\tosc} \approx
  \frac{\calG_\calE f_\mathrm{lw}}{\tprec}\,\CoefK\calRsep\sqrt{\calRcapt}.
\end{equation}

In the opposite limit, a star that achieves $\calR<\calRcapt$ while being far from 
the \sbh\ may precess out of the loss cone before reaching periapsis, similar 
to what happens in the full-loss-cone case of the spherical problem. 
Then the capture rate is less than given by the above equation, because not all stars 
in the loss wedge are captured after one precession period. It is easy to see that 
in this case, which can be called the ``full loss wedge regime'', the rate of consumption 
is equivalent to the draining rate of the full loss cone (\ref{eq_Fdrain}).
In other words, the precession is fast and shuffles stars 
in angular momentum quickly enough that the loss cone stays full, 
hence the capture rate is just the instantaneous number of stars inside the loss cone 
divided by their radial period. 
By ignoring the effects of a finite precessional time,
\citet{MagorrianTremaine1999} were essentially in this regime.

By analogy with the spherical case, we introduce the quantity $q_\mathrm{axi}$ 
separating the two regimes:\footnote{\citet{MerrittVasiliev2011} defined an analogous
quantity for pyramid orbits in the triaxial geometry, their equation (55).}
\begin{equation}   \label{eq_q_axi}
q_\mathrm{axi} \equiv \frac{\calF_\mathrm{drain,lw}}{\calFdrain} 
  = \frac{\trad}{\tosc} \sqrt{\frac{\calRsep}{\calRcapt}}
  = \frac{\trad}{\tprec}\frac{\CoefK \calRsep}{\sqrt{\calRcapt}}.
\end{equation}
It is easy to see that $q_\mathrm{axi} \gg 1$ at the radius of influence, 
where $\trad \approx \tprec$, as long as the flattening is not too small 
($\calRsep \gg \sqrt{\calRcapt}$). Unlike the spherical problem, the transition from 
empty- to full-loss-wedge regimes always occurs well within the radius of influence, 
and therefore the main contribution to the total capture rate comes from the 
full-loss-wedge regime. Moreover, for most realistic cases $q_\mathrm{axi} \gg q$ for 
the entire range of radii. Indeed, combining expresssions
(\ref{eq_Tprec}, \ref{eq_q_spher}, \ref{eq_Trel}, \ref{eq_q_axi}) and substituting 
$\calRcapt=2r_\mathrm{lc}/a$, where $a$ is the orbit semimajor axis, we obtain
\begin{subequations}
\begin{equation}
\frac{q_\mathrm{axi}}{q} \approx \frac{\sqrt{\calRsep\calRcapt}}{\Dif\tosc} 
  \approx \frac{\calRsep}{5\ln\Lambda} \frac{\Mbh}{m_\star} \sqrt{\frac{r_\mathrm{lc}}{a}}
\end{equation}
This ratio decreases with radius; evaluating it at $a=\rh$ and substituting 
$r_\mathrm{lc} \ge 8\rg = 8G\Mbh/c^2$ (see \S~\ref{sec_estimates}), 
$\rh \approx G\Mbh/\sigma^2$, where $\sigma$ is the velocity dispersion of stars 
outside $\rh$, and $\ln\Lambda\sim 20$, we may rewrite the above expression as
\begin{equation}
\frac{q_\mathrm{axi}}{q} \approx 
  \frac{\calRsep}{0.1} \frac{\Mbh}{10^6\,M_\odot} \frac{\sigma}{100\,\mathrm{km\ s}^{-1}} \,,
\end{equation}
\end{subequations}
which is likely to be $\gtrsim 1$ if the flattening is not too small.
In other words: changes in angular momentum near the loss cone 
boundary are determined by the regular precession ($q_\mathrm{axi}$), 
and not by relaxation ($q$).

To summarize, the boundary condition in the axisymmetric problem is
\begin{equation}  \label{eq_Flux_axi_combined}
\calF_\mathrm{lw} \approx \calFdrain\,\mathrm{min}(1, q_\mathrm{axi}) \;,
\end{equation}
and the relation between the flux and the value of $f$ at the boundary, 
expressed in the same way as in the spherical problem (\ref{eq_boundary_cond_spher}), reads
\begin{subequations}
\begin{eqnarray}  \label{eq_boundary_cond_axi}
\calG_\calE f_\mathrm{lw} &=& \alpha_\mathrm{axi}\Dif^{-1}\calF_\mathrm{lw}, \\
\alpha_\mathrm{axi} &=& 
\begin{cases} 
  q/q_\mathrm{axi} &\mbox{if } q_\mathrm{axi} < 1, \\
  q & \mbox{if } q_\mathrm{axi} > 1. 
\end{cases}  \label{eq_alpha_axi}
\end{eqnarray}
\end{subequations}

The derivation above only gives this relation in ``integrated'' form, that is, 
one coefficient $\alpha_\mathrm{axi}$ for the entire $\calH-\calR_z$ plane at 
a given energy.
While this is certainly an oversimplification, we argue below that it does not
greatly affect the capture rate.

The distinction between empty and full loss cones in the spherical problem 
depends on whether $\calF_\mathrm{lc} \ll \calFdrain$ 
(equation \ref{eq_Flux_sph}) or not, 
or equivalently whether $\alpha \gg q$. 
In the spherical case,  $\alpha \approx q\,\mathrm{max}(1,q^{-1/2})$, 
and the transition occurs at $q\approx 1$.
In the axisymmetric problem, the distinction between empty and full loss {\it wedges} 
is  whether $\calF_\mathrm{lc} \ll \calFdrain$ or not, and 
the transition is at $q_\mathrm{axi} = 1$.
In most realistic cases, $q_\mathrm{axi} \gg q$, although it is
not necessarily true that $q_\mathrm{axi}>q^{1/2}$.
Therefore, the coefficient $\alpha_\mathrm{axi}$ in the boundary condition
(\ref{eq_boundary_cond_axi}) can be both greater or less than its spherical 
counterpart $\alpha$ (equation \ref{eq_alpha_q}) for the same $\calE$ and $\calRcapt$.
In the case $q>1$ (full loss cone of the spherical problem) there is essentially 
no difference in boundary conditions since $q_\mathrm{axi}$ is also greater than $1$. 
In the opposite case ($q<1$), $\alpha_\mathrm{axi}<1$ regardless of the value of $q_\mathrm{axi}$, 
and the capture rate turns out to depend only weakly on it, as argued in the next section.
In this latter case $q_\mathrm{axi}$ may be both greater or less than unity, i.e. 
the loss wedge may be either empty or full. 
However, as we noted above, the relation between the capture rate 
and $\overline f$ in the axisymmetric problem depends not only on the boundary condition, 
but also on the structure of the entire solution, as addressed in the next section. 

\section{Solution of the two-dimensional Fokker-Planck equation}   \label{sec_FP_solution}

We are  primarily interested here in the capture rate, i.e. the 
flux of stars into the \sbh, and not in the evolution of the mass distribution
(density profile, flattening) which we assume to be fixed.
The flux is determined mainly by diffusion in angular momentum.
Accordingly,  we consider the two-dimensional Fokker-Planck equation 
describing evolution in ($\calH, \calR_z$) and neglect diffusion 
in energy.
We note that most of our results are quoted for a  density profile 
$\rho\propto r^{-3/2}$ which is reasonably close to the 
Bahcall-Wolf stationary solution, $\rho\propto r^{-7/4}$, further
justifying the neglect of energy evolution.

\subsection{Previous studies}

To date, all studies of axisymmetric systems assumed that the distribution function depends 
only on the two classical integrals of motion, $\calE$ and $\calR_z$.
Then it is easy to show that $f(\calR_z) \propto \sqrt{\calR_z}+\mathrm{const}$ for $\calR_z\ll 1$.
Indeed, from (\ref{eq_difcoefRz2}) we see that $\langle(\Delta\calR_z)^2\rangle \propto \calR_z$
for small $\calR_z$, and from (\ref{eq_densityofstates}) that the density of states 
$\calGav\propto \calR_z^{-1/2}$.
Then the total capture rate per unit energy is 
$\calF \propto \calGav\,\langle(\Delta\calR_z)^2\rangle\, \d f/\d\calR_z$,
and should be independent of $\calR_z$, which leads to the square-root profile of $f(\calR_z)$.

\citet{MagorrianTremaine1999} used this argument to derive the relation between 
the capture rate $\calF$ and the average value of the distribution function at a given energy 
$\overline f$, as follows.
Start by writing the relation between $\calF$ and the value of $f$ at the 
loss wedge boundary $f_\mathrm{lw} \equiv f(\calR_z=\calRcapt)$, which corresponds 
essentially to the full loss wedge regime (equation~\ref{eq_boundary_cond_axi} with 
$q_\mathrm{axi}=1$). 
Then express the integrated flux in the $\calR_z$ direction as
\begin{equation}  \label{eq_FluxRz}
\calF = B\,\Dif\calG_\calE \sqrt{\calR_z} \frac{\d f}{\d\calR_z} \;,
\end{equation}
independent of $\calR_z$.
The numerical factor $B$ is related to the ``area of the loss wedge'', $B\approx\sqrt{\calR_m}/\pi$, 
where $\calR_m$ is the peak angular momentum of saucer orbits and may be associated 
with our definition of $\calRsep$. The distribution function is then
\begin{equation}  \label{eq_f_Rz}
f(\calR_z) = f_\mathrm{lw} + \frac{2\calF}{B\,\Dif\calG_\calE} \sqrt{\calR_z} =
  f_\mathrm{lw} \left(1 + \frac{2}{Bq}\sqrt{\calR_z}\right) .
\end{equation}
The average distribution function is
\begin{equation}  \label{eq_f_avg_over_Rz}
\overline f(\calE) = \int_0^1 d\calR_z\,w(\calR_z)f(\calR_z) \;,
\end{equation}
where $w(\calR_z) d\calR_z$ is the fraction of the phase space volume at a given $\calR_z$.
\citet{MagorrianTremaine1999} took $w=1/(2\sqrt{\calR_z})$;
a more correct value is $w=1/\sqrt{\calR_z}-1$.
Using their value, one finds 
$$
\overline f = f_\mathrm{lw} \left( 1 + \frac{1}{2} \frac{2}{Bq}\right) \;;
$$
the correct expression would contain $1/3$ instead of $1/2$ in the brackets.
The relation between the steady-state capture rate per unit energy and the average (isotropized) 
value of the distribution function  is
\begin{equation}  \label{eq_fluxMT}
\calF_\mathrm{MT} = \frac{\Dif\,\overline \calN}{q + \pi/\sqrt{\calRsep}} 
\end{equation}
where $\overline\calN \equiv \calG_\calE \overline f$.

Comparing of this expression with the analogous one in the spherical case 
(\ref{eq_capt_rate_spher}), we see that when $q \gg 1$, the capture rate is 
essentially the same as in the spherical case (full-loss-cone regime), while 
in the opposite limit it is determined by the diffusion coefficient $\Dif$ 
and the value of  $\calRsep$, rather than by the size of the loss cone $\calRcapt$.
This is a consequence of the geometry of loss wedge boundary, which stretches 
in one direction to a fixed fraction $\calRsep$ of the phase space.

\subsection{The present study}

\begin{figure}
$$\includegraphics{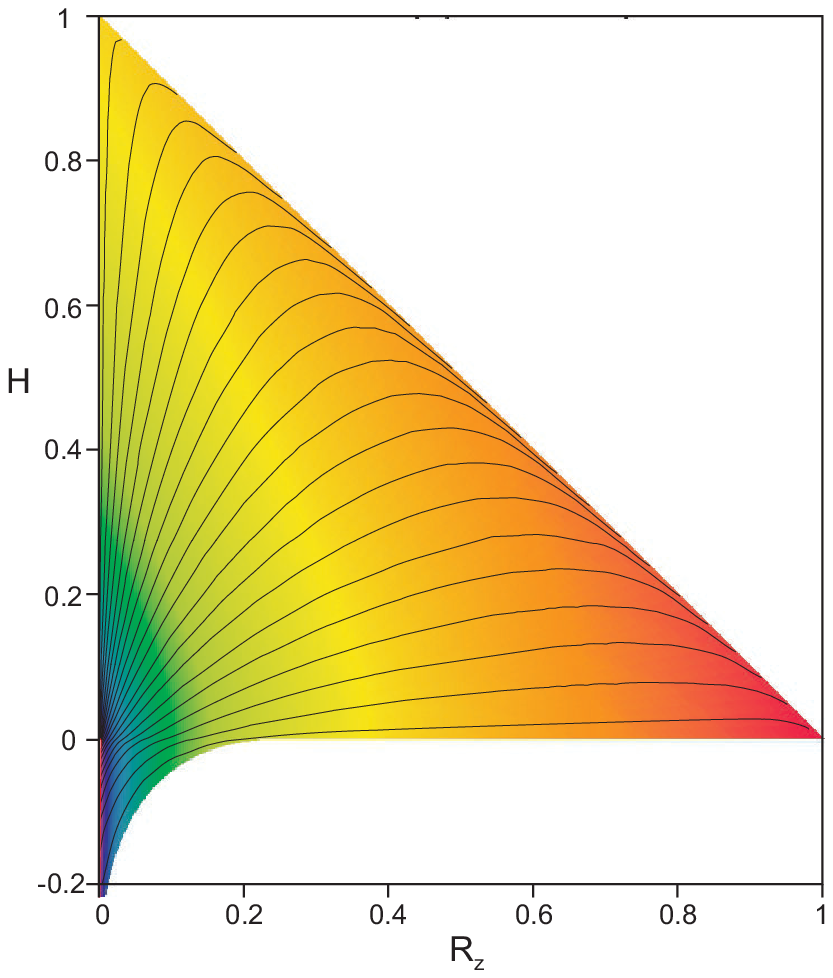}$$
\caption{($\calR_z,\calH$) phase plane showing stream lines
in a quasi-stationary solution to the two-dimensional diffusion problem, 
for $\calRsep=0.25$ and $\calRcapt=0.003$. 
The capture boundary $\calRmin(\calH,\calR_z)$ (red) stretches from 
$\calH\approx -\calRsep$ to $\calH=\calRcapt$ almost parallel to the ordinate  
(for an exagerrated close-up refer to Figure~\ref{fig_phaseplane} 
where lines of constant $\calRmin$ are shown in solid black).
More than one-half of the flux lines end up in the saucer region (at $\calH<0$). 
Fill color shows the value of $f$.
} \label{fig_streamlines}
\end{figure}

\begin{figure}
$$\includegraphics[angle=-90]{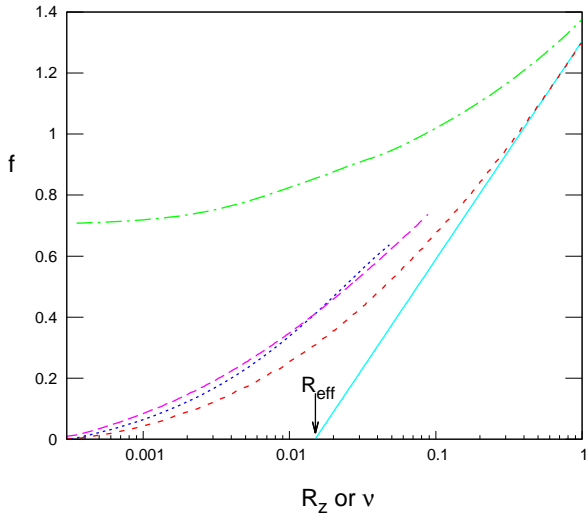}$$
\caption{
Steady-state numerical solution of 2d problem for 
$\calRsep=0.1, \calRcapt=3\times 10^{-4}, \alpha=0$, together with asymptotical profiles 
at small $\calR_z$ and large $\nu\equiv \calH+\calR_z$.
Red short-dashed line is $f(\nu)$ in the tube region, averaged over lines of constant 
$\nu$ (which roughly correspond to average $\calR$ for $\nu\gtrsim \calRsep$);
purple long-dashed line is $f(\calR_z)$ in the saucer region, averaged over lines 
of constant $\calR_z$.
Blue dotted line is the approximation (\ref{eq_f_Rz_saucer}) for the saucer region, 
and solid cyan line is the approximation (\ref{eq_f_R_tube}) for the main (tube) region
of phase space, which intersects the abscissa axis at the effective capture boundary 
$\calReff \approx 0.015$ (\ref{eq_Reff_axi}).
These two asymptotical profiles do not intersect at $(\calRsep,f_\mathrm{sep})$ 
as they would if we used the crude approximation for $\calReff$ described in the text;
the actual value for $\calReff$ was computed from the numerical 2d solution and is 
roughly twice as higher than the simple approximation. 
Green dot-dashed line shows the value of $f(\calR_z)$ for the entire phase space, 
averaged over the second coordinate; it is not at all close to square-root profile 
used by \citet{MagorrianTremaine1999} and does not tend to zero at $\calR_z=\calRcapt$, 
since most part of phase space at fixed $\calR_z$ is occupied by tube orbits which 
do not come close to the capture boundary.
} \label{fig_Rprofile}
\end{figure}

\begin{figure*}
$$\includegraphics[angle=-90]{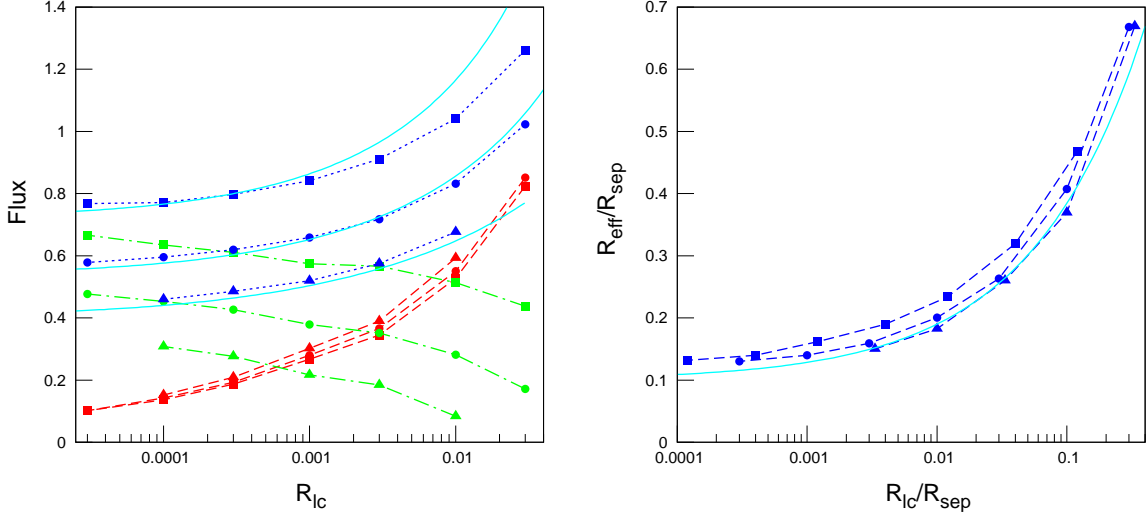}$$
\caption{
Stationary fluxes and effective capture boundaries for a series of 2d problems 
with $\alpha=0$ (empty-loss-cone limit). 
Top curves (boxes) have $\calRsep=0.25$; 
middle curves (circles) are for $\calRsep=0.1$; 
bottom curves (triangles) have $\calRsep=0.03$. 
The left panel shows the ratio of the stationary capture rate to the
 average value of distribution  function, $\calF/\overline f$. 
Red dashed curves are the capture rate from tube orbits, 
green dot-dashed curves from saucer orbits, and blue dotted curves are the total capture rate.
The blue solid curves show the approximation of equation (\ref{eq_capt_rate_axi}). 
The right panel plots the effective capture boundary $\calReff$ normalized to $\calRsep$.
Blue dashed lines are results fitted from the 2d solution (best-fit logarithmic 
profile for $\calR>\calRsep$, like the green dot-dashed line in 
Figure~\ref{fig_Rprofile}); blue solid curves are the approximation of equation
(\ref{eq_Reff_axi}). 
} \label{fig_Reff}
\end{figure*}

We set up an initially uniform (isotropic) distribution ($f=\const$) in the $\calH-\calR_z$
plane outside the capture boundary, defined by equating $\calRcapt$ and $\calRmin$ 
found from equation(\ref{eq_Rminmax}) ($\calR_2$ for saucers and $\calR_3$ for tubes); 
a series of isolines of constant $\calRmin$ is plotted in Figure~\ref{fig_phaseplane}. 
We studied a range of values for both $\calRsep$ and $\calRcapt \ll \calRsep$, 
as well as various parameters $\alpha_\mathrm{axi}$ in the boundary condition (\ref{eq_alpha_axi}).

The numerical solution of equation (\ref{eq_FPflux}) was obtained on a non-uniform rectangular 
grid using two different sets of coordinates, defined such that the capture boundaries are 
parallel to the coordinate axes (see Appendix \ref{sec_appendix_coords} for details);
grid sizes were typically $100-300$ in each dimension and the loss region was resolved 
by 10-20\% of the grid cells.
We advanced the solutions until time $T=1/\Dif$ to achieve a steady-state profile, 
from which we could extract the relation between the capture rate and the average value of $f$.

In the spherical case, the solution is controlled by two parameters 
(aside from $\Dif$ which scales the time): 
the capture boundary $\calRcapt$ and the boundary coefficient $\alpha$ (or $q$).
As regards the steady-state profile, these two parameters are not independent, 
since one can always transform the problem to another (primed) one with 
$\alpha'=0$ and $\calRcapt'=\calR_0$ (equation~\ref{eq_R0}), 
so the family of solution is effectively one-parametric. 
To compare the time-dependent solution of the axisymmetric problem to the spherical case,
we introduce the concept of ``equivalent spherical problem'', that is, the one-dimensional 
problem with the same coefficient $\alpha$ in the boundary condition as $\alpha_\mathrm{axi}$ 
in equation~(\ref{eq_alpha_axi}), and with some effective capture boundary $\calReff$ 
chosen such that the time-dependend capture rate closely follows that of the axisymmetric 
problem. Our goal is then to find $\calReff$ as a function of the loss cone size
$\calRcapt$ and degree of flattening, the latter parametrized by $\calRsep$.

In the remainder of this section we present simple analytical arguments that 
give a qualitatively correct description of the two-dimensional numerical solution 
of the axisymmetric problem, and provide a fitting formula for $\calReff$.

Figure~\ref{fig_streamlines} shows stream lines of flux and isocontours of $f$ 
in a quasi-stationary 2d solution for a rather exaggerated value of 
$\calRcapt=0.003 \approx 10^{-2}\calRsep$. 
Even in this case, most of the stream lines 
end inside the saucer region, and that is definitely so for more realistic (smaller) 
values of $\calRcapt$. 
It is also clear that in the saucer region, $f$ depends mainly on $\calR_z$ and 
is almost independent of the second coordinate, which justifies the square-root 
profile of $f(\calR_z)$ as in equation~(\ref{eq_f_Rz}), but only in this region.
In the tube region, for the greater part of the phase space ($\nu \equiv 
\calH+\calR_z \gtrsim \calRsep$), the solution is close to that of the spherical problem, 
that is, $f(\calR) \propto \ln\calR + \const$, with $\calR\approx\nu$ experiencing only 
small oscillations. 
We can build an approximate solution by joining the two asymptotic forms at $\calRsep$.
A better description for the saucer region accounts for the fact that the flux in the 
$\calR_z$ direction gradually decreases from $\calF$ at the capture boundary to zero at $\calRsep$:
\begin{equation}  \label{eq_FluxRz_corr}
\calF_{\calR_z} \equiv \int_{\calH_\mathrm{min}}^0 \!\!\!\!\! d\calH \, 
  \frac{1}{2}\langle(\Delta\calR_z)^2\rangle \calGav \,\frac{\d f}{\d \calR_z} 
\approx \calF \times \left(1-\sqrt{\frac{\calR_z}{\calRsep}}\right) .
\end{equation}
$\calH_\mathrm{min}(\calR_z)$ is the minimum value of $\calH$, corresponding to 
the fixed-point saucer orbit (\ref{eq_FPO}). 
The second, approximate equality in equation (\ref{eq_FluxRz_corr}) 
is an empirical fit to the numerical 2d solution.
Using the asymptotic expressions (\ref{eq_coef_asympt_SAU}), it is easy to show that
for $\calR_z\ll \calRsep$ the flux has the form (\ref{eq_FluxRz}), with the numerical 
coefficient $B \approx 0.4\sqrt{\calRsep}$ (which is $\sim 30\%$ larger than the value 
used by \citet{MagorrianTremaine1999}). 
The solution in the saucer region is obtained 
by solving the differential equation (\ref{eq_FluxRz_corr}):
\begin{equation}  \label{eq_f_Rz_saucer}
f_\mathrm{saucer} (\calR_z) = f_\mathrm{lw} + \frac{2\calF}{B\,\Dif\calG_\calE} 
  \sqrt{\calR_z} \left(1 - \frac{1}{2} \sqrt{\frac{\calR_z}{\calRsep}} \right) ,
\end{equation}
which is a somewhat improved form of equation~(\ref{eq_f_Rz}).
The solution in the tube region outside $\calRsep$ is approximated by 
\begin{equation}  \label{eq_f_R_tube}
f_\mathrm{tube} (\calR) = f_\mathrm{sep} + 
  \frac{\calF}{\Dif\calG_\calE} \ln\frac{\calR}{\calRsep} \;;
\end{equation}
the coefficient of the logarithmic term gives the same flux in the $\calR$ direction as in the 
spherical problem, and $f_\mathrm{sep} \equiv f_\mathrm{saucer}(\calRsep)$.
Figure~\ref{fig_Rprofile} shows the two asymptotic expressions along with the actual numerical solution.

We compute the isotropized value $\overline f$ taking into account only the contribution 
from $f_\mathrm{tube}$, which introduces a fractional error of at most $\calRsep$:
\begin{equation}
\overline f \approx \int_{\calRsep}^1 f_\mathrm{tube}(\calR)\, d\calR \approx 
  f_\mathrm{sep} + \frac{\calF}{\Dif\calG_\calE} \left(\ln\frac{1}{\calRsep}-1\right) .
\end{equation}

Putting all this together and expressing the relation between $\calF$ and $\overline \calN$ 
in terms of the coefficient $\alpha_\mathrm{axi}$ (\ref{eq_boundary_cond_axi}), we obtain
\begin{equation}  \label{eq_capt_rate_axi_approx}
\calF = \frac{\Dif\,\overline \calN}
  {\alpha_\mathrm{axi} + \ln(1/\calRsep) - 1 + 2\sqrt{\calRsep}/B} .
\end{equation}
Equation~(\ref{eq_capt_rate_axi_approx}) can be compared with 
equation~(\ref{eq_fluxMT}) of \citet{MagorrianTremaine1999}:
both share the property of being independent of $\calRcapt$, replacing it with
some effective capture boundary $\calReff$ for the empty-loss-cone regime, 
although this effective value is different. 
By comparing (\ref{eq_capt_rate_axi_approx}) with the spherical analog (\ref{eq_capt_rate_spher}),
we see that in our approximation, $\calReff = \calRsep\exp(-1/B) \approx 0.08\calRsep$.
Figure~\ref{fig_Reff} shows that equation~(\ref{eq_capt_rate_axi_approx}) predicts well 
the flux in the numerical steady-state solutions; a better approximation to the effective capture 
boundary and the capture rate is 
\begin{eqnarray}  \label{eq_Reff_axi}
\calReff &=& \calRsep\, \left(0.1 + 0.9\sqrt{\frac{\calRcapt}{\calRsep}}\right) , \\
\calF &=& \frac{\Dif\,\overline \calN}
  {\alpha_\mathrm{axi} + \ln(1/\calReff) - 1} . \label{eq_capt_rate_axi}
\end{eqnarray}

Overall, the capture rates in the axisymmetric geometry are higher than in the 
spherical case with the same boundary condition $\alpha=\alpha_\mathrm{axi}$, 
but not by a large factor: 
in the full-loss-cone regime ($\alpha\gg 1$) they are  essentially the same, 
while in the empty-loss-cone regime the effective boundary $\calReff$ is higher than 
$\calRcapt$, but since the flux depends on it only logarithmically, the difference 
is not likely to be more than a factor of a few. 

Of course, more relevant is a comparison that takes into account that $\alpha$ 
in the spherical case may be different from $\alpha_\mathrm{axi}$ for the same values 
of $\calE$ and $\calRcapt$, as noted near the end of the previous section.
For the least bound stars which are in the full-loss-cone regime ($q(\calE) \gg 1$), 
$q_\mathrm{axi} \gg 1$ and $\alpha_\mathrm{axi} \approx \alpha \approx q$. 
In this case, the capture rate does not depend on the diffusion coefficient 
$\Dif(\calE)$ but only on the value of $\calRcapt$. 
The boundary condition (\ref{eq_boundary_cond_axi}) states that the flux $\calF_\mathrm{lw}$ 
is proportional to the average, isotropized value $\overline f\approx f_\mathrm{lw}$, 
and $\Dif$ cancels out.

In the opposite case $q(\calE) \lesssim 1$, the capture rate is limited by diffusion, and  
$\tilde f(\calR)$ is no longer close to isotropic.
If $q \lesssim 1$, $\alpha_\mathrm{axi}$ is also $\lesssim 1$, and the denominator 
in the expression for the capture rate (\ref{eq_capt_rate_axi}) tends to some constant value 
which depends only on $\calRsep$, but not on $q$ or $\calRcapt$ 
(provided that $\calRcapt \ll \calRsep$).
It is largely irrelevant whether the boundary condition itself corresponds to 
the empty ($q_\mathrm{axi}<1$) or full-loss-wedge regime. 
In other words, in this \textit{diffusion-limited regime} (both in spherical and 
axisymmetric cases) the boundary conditions (\ref{eq_boundary_cond_spher}) or 
(\ref{eq_boundary_cond_axi}) determine $f_\mathrm{lc}$ for a given flux $\calF_\mathrm{lc}$, 
which itself is set by the gradient of the overall steady-state profile of solution.
By comparison, in the spherical case the denominator in the expression for 
the capture rate (\ref{eq_capt_rate_spher}) also depends only weakly (logarithmically) 
on $\calRcapt$ and is almost independent of $\alpha$. 
Therefore, the difference in capture rates between the axisymmetric and spherical problems, 
which results from the difference between $\ln\calReff(\calRsep,\calRcapt)$ and $\ln\calRcapt$,
is at most a factor of few in the case $q(\calE) \lesssim 1$.

\section{The role of chaotic orbits}  \label{sec_chaos}

The Fokker-Planck formalism developed in the previous sections relied on the existence 
of three integrals of motion deep inside the \sbh\ influence region.
Apart from some special, fully integrable cases 
\cite[e.g.][]{SridharTouma1997}, most axisymmetric potentials containing central 
point masses are characterized by chaotic motion in the low-angular-momentum parts of 
phase space beyond the influence radius. 
Chaotic orbits still respect two integrals of the motion, $\calE$ and $\calR_z$, 
but in the absence of a third integral, they can in principle fill the accessible region 
in the meridional plane, allowing them to be captured as long as
$\calR_z<\calRcapt$.
We estimate the capture rate from these orbits by the following argument, 
similar to an argument of \citet{MagorrianTremaine1999}. 
First we introduce the concept of draining of the loss region, arising from non-conservaton 
of angular momentum without any relaxation effects, then estimate the role of relaxation, 
and finally discuss the combined effects of draining and relaxation.

Assume that the chaotic orbits occupy a region in the $\calR-\calR_z$ 
plane with $\calR<\calRch$. 
The value of $\calRch$ plays the same role as $\calRsep$ inside the radius of influence, 
and is comparable to it for the same degree of flattening.
Furthermore, we assume that every chaotic orbit with a given $\calR_z$ can attain values 
of $\calR\in [\calR_z\ldots\calRsep]$ with equal probability (numerical tests verify that 
this is a reasonable assumption). 
Recall that the number of stars with given $\{\calR,\calR_z\}$, which may be identified 
with the probability of finding a star in a given interval of $d\calR d\calR_z$, 
is $d\calN = \calG_\mathrm{sph}\,f(\calR,\calR_z)\,d\calR\,d\calR_z$, with 
$\calG_\mathrm{sph}$ given by equation~(\ref{eq_densityofstates}).
Then the fraction of time such an orbit spends below the capture boundary is 
$(\sqrt{\calRcapt}-\sqrt{\calR_z})/\sqrt{\calRch}$, and this is essentially the 
probability of being captured during one radial period 
(assuming the full-loss-cone boundary condition, i.e. 
that the change in angular momentum during one  period is much 
larger than the capture boundary, which is reasonable for chaotic orbits). 

Next we evaluate the time-dependent rate of capture of stars from chaotic orbits.
From the above argument it follows that we need to consider $f(\calR_z)$ decaying 
exponentially at every value of $\calR_z$ from its initial value $f_\mathrm{init}$, 
but with a different rate:
$$
f(\calR_z,t) = f_\mathrm{init}\,\exp\left[-\frac{t}{\trad}
  \frac{\sqrt{\calRcapt}-\sqrt{\calR_z}}{\sqrt{\calRch}}\right] .
$$
The total number of chaotic orbits with $\calR_z<\calRcapt$ and 
their capture rate is then given by
\begin{subequations}
\begin{eqnarray}
N_\mathrm{ch}(\calE,t)\,d\calE &=& 
  \int_0^{\calRcapt} \calG_\calE f(\calR_z,t) 
  \left(\sqrt{\calRch/\calR_z}-1\right) d\calR_z\, d\calE  \nonumber\\
  &\approx& \calG_\calE f_\mathrm{init}\, 2\sqrt{\calRch\calRcapt}\,
  \frac{1-\exp(-2\tau)}{2\tau} \,d\calE  , \label{eq_Nchaotic} \\
\calF_\mathrm{ch}(\calE,t)\,d\calE &=&
  \frac{\calG_\calE f_\mathrm{init} \calRcapt}{\trad} \,
  \frac{1-(2\tau+1)\exp(-2\tau)}{2\tau^2} \,d\calE ,  \nonumber \\
\label{eq_Fchaotic}
\end{eqnarray}
\end{subequations}
where
\begin{equation}  \label{eq_Tdrainch}
\tau \equiv t/T_\mathrm{drain} \;,\quad 
  T_\mathrm{drain} \equiv 2\trad\sqrt{\calRch/\calRcapt}  .
\end{equation}
From here it is clear that if we identify $f_\mathrm{init}$ with the initial value 
of the distribution function in the loss cone $f_\mathrm{lc}$, then the capture rate is 
initially equal to the draining rate of a uniformly populated loss cone (\ref{eq_Fdrain}).
In particular, when $f_\mathrm{init}=\overline f$, we recover the standard, full-loss-cone 
draining rate, regardless of the value of $\calRch$. On the other hand, the draining time 
does depend on $\calRch$, since the number of stars in the chaotic region to be drained 
is $2\sqrt{\calRch/\calRcapt}$ times larger than the number of stars in the loss cone, 
therefore the draining time is longer than the radial period by the same factor.
At times much longer than the draining time, the capture rate declines as $t^{-2}$,
not exponentially,
since it is dominated by the draining of chaotic orbits with 
$\calRcapt-\calR_z \ll \calRcapt$.

There is a great deal of similarity between the capture rates from the loss wedge 
of the saucer region of phase space for regular orbits within the radius of influence, 
and the chaotic region outside it. 
The details of draining are somewhat different (in particular, for the regular orbits 
the draining rate declines as $t^{-3}$, as noted by \citet{MagorrianTremaine1999}), 
but since the draining time for saucer orbits is usually much shorter than a Hubble time,
we ignore that distinction and adopt the same expressions for them as for chaotic orbits.
In both cases, the \textit{local} boundary condition for the loss region corresponds 
to the full loss cone (\ref{eq_boundary_cond_axi}), at least for the case 
$q_\mathrm{axi}>1$ relevant for all but the most tightly bound orbits. 
However, the \textit{global} shape of the steady-state solution depends on 
whether the overall flux into the low angular momentum region is limited by 
diffusion ($q(\calE)\lesssim 1$) or not. In the first case, the steady-state 
solution will still have a logarithmic form for $\calR\gtrsim \calRch$ or $\calRsep$, 
corresponding to some effective capture boundary $\calReff$, and the capture rate 
depends on this effective boundary only logarithmically. In the latter case, 
the capture rate is essentially the full loss cone rate for an isotropic distribution 
function. The latter case, however, is rarely attained because $q(\calE)$ rapidly 
drops with decreasing binding energy. On the other hand, if the draining time for 
chaotic orbits is comparable to the Hubble time, then their capture rate may 
still be quite high even in the absence of relaxation, provided that the initial value 
$f_\mathrm{init}$ of the distribution function inside the chaotic region was not 
much different from the isotropic value $\overline f$.

The combined effect of draining and relaxation may be approximately accounted for 
by the following recipe. Let $\calF_\mathrm{rel}(\calE, t) \,d\calE$ be the capture 
rate per unit energy from the Fokker-Planck equation with initial conditions 
corresponding to the loss region being initially empty (i.e. the solution considered 
in \S~\ref{sec_FP_solution}). Since the loss region initially may have some 
nonzero value of $f$, $0\le f_\mathrm{init} \le \overline f$, 
the phase-space gradient of $f$ near the loss region boundary will be less than 
arising in the Fokker-Planck solution, and the capture rate from relaxation alone 
may also be lower. 
We approximate the total capture rate by the sum of the draining rate $\calF_\mathrm{drain}$ 
and the collisional flux $\calF_\mathrm{rel}$ multiplied by 
$1-N_\mathrm{ch}(t)/N_\mathrm{ch,0}$, where $N_\mathrm{ch,0}$ is the number of 
chaotic orbits at $t=0$ with $f_\mathrm{init}=\overline f$.
This expression is used to compare Fokker-Planck models against $N$-body simulations
in \S\ref{sec_Nbody} and to compute the capture rates for real galaxies in 
\S\ref{sec_estimates}.
It is important to note that the effective capture boundary $\calReff$ defined in 
(\ref{eq_Reff_axi}), as the parameter controlling the overall shape of the steady-state 
solution and the gradient of the distribution function (and hence the capture rate due 
to diffusion), is not the same as the size of the loss region 
($\sqrt{\calRsep\calRcapt}$ for saucer orbits, $2\sqrt{\calRch\calRcapt}$ for chaotic 
orbits) which determines the draining time. The former, being a fixed fraction of 
$\calRsep$ ($\calRch$), is usually much larger than the latter.

These results will be used in \S \ref{sec_Nbody} when we compare the model
predictions with the results of $N$-body simulations.

\section{Triaxiality}  \label{sec_triaxiality}

For completeness, and to put our results in a broader context, we briefly discuss 
the case when the stellar cusp around the \sbh\ is triaxial. 
Triaxial potentials support two distinct familes of tube orbits, circulating about the
long and short axes of the triaxial figure \citep{MerrittVasiliev2011}.
In addition, there is a new class of centrophilic regular orbits, the 
pyramids \citep[][Figure 11]{MerrittValluri1999}. 
The defining feature of  pyramids is that $\calRmin=0$ for all of 
them,\footnote{Relativistic precession alters this conclusion for the most bound orbits 
\citep{MerrittVasiliev2011}.}
and a star on such an orbit will eventually find its way into the \sbh\ even without 
the assistance of collisional relaxation. 
The fraction of phase space occupied by pyramids is comparable to that of saucer orbits,
i.e. $\sim \calRsep$.
Outside the radius of influence, the regular pyramid orbits are mostly replaced by chaotic 
orbits, which are however still centrophilic \citep{PoonMerritt2001}.

\begin{table}
\caption{Comparison of three geometries\label{tab_geoms}}
\begin{tabular}{lccc}
\hline
&Spherical & Axisymmetric & Triaxial \\
Fraction of stars\\ with $\calRmin<\calRcapt$ & 
$\calRcapt$ & $\sqrt{\calRcapt\calRsep}$ & $\calRsep$ \\
Draining time $T_\mathrm{drain}$ &
$\trad$ & $\gtrsim \tosc$ & $\gg \tosc$ \\
\hline
\end{tabular} 
\end{table}

Table~\ref{tab_geoms} summarizes the three cases.
The number of stars that potentially can be captured (loss region, stars with 
$\calRmin<\calRcapt$) increases with decreasing symmetry, however the instantaneous 
number of stars in the loss cone is the same (the fraction of time a star on a loss
region orbit actually has instantaneous $\calR<\calRcapt$ exactly balances that).
Consequently, the survival time of these stars also increases with decreasing symmetry:
$T_\mathrm{drain}$ is larger than $\trad$ by the same factor as the number of stars 
in the loss region versus the loss cone, assuming a full-loss-cone draining rate.

In the absence of relaxation, the loss region is rapidly depleted
in both spherical and axisymmetric cases (the latter -- except for the most massive 
\sbhs), but in the triaxial case the draining time 
of the loss region may be comparable to or even exceed galaxy lifetimes 
(\citet{MerrittVasiliev2011}; in that paper a rather small departure from spherical symmetry 
was considered; for $\calRsep\sim 0.1$ or for more massive \sbhs\ the time will be  longer.)
If, as suggested by \citet{MerrittPoon2004} and \citet{HolleySigurdsson2006}, 
the capture of stars from centrophilic orbits (both regular pyramids inside the radius 
of influence, and chaotic orbits outside it) can sustain a full-loss-cone feeding rate for 
an initially isotropic distribution of stars, then it will remain near this level for a time
$\sim T_\mathrm{drain}$. For stellar systems older than that, it is necessary to take 
2-body relaxation into account, and the outcome will probably be similar to what 
happens in the $q_\mathrm{axi}>1, q<1$ regime of the axisymmetric problem: reshuffling 
of stars in angular momentum near the loss cone boundary due to nonspherical torques is 
efficient enough to keep the loss cone full, but the value of $f_\mathrm{lc}$ near loss cone 
is much smaller than the average (isotropized) value $\overline f$, because the supply 
of stars into the low-$\calR$ region is limited by the diffusion from higher $\calR$.
Extrapolating the estimates of draining times for axisymmetric galactic models from 
\S \ref{sec_estimates} to the triaxial case, one may conclude that for most massive galaxies 
the lifetime of centrophilic orbits may indeed be longer than Hubble time, provided that 
the triaxiality is not destroyed by the effects of chaos \citep{MerrittQuinlan1998}.

We stress that  genuinely centrophilic orbits exist only in the triaxial case, since 
in the axisymmetric geometry the conservation of $L_z$ precludes orbits from reaching 
arbitrarily small radii.
However, some degree of non-axisymmetry is to be expected
in every real galaxy.

\section{Comparison with $N$-body simulations}   \label{sec_Nbody}

To test the predictions of the Fokker-Planck models, we carried out a series of $N$-body 
integrations of both spherical and flattened models of galaxies containing
central point masses.
The model mass distribution was a flattened modification of the
spherical \citet{Dehnen1993} profile:
\begin{equation}  \nonumber
\rho(\boldsymbol{x}) = \frac{4\pi}{(3-\gamma)} \frac{1}{r^{\gamma}(1+r)^{4-\gamma}} 
  \left(1+\epsilond\left[\frac{z^2}{r^2}-\frac{1}{3}\right]\right) ,
\end{equation}
with $\gamma=3/2$. 
This model deviates from the scale-free profile of equation~(\ref{eq_rho_cusp})
at large radii but is close to it inside $r_\mathrm{infl}$.

Our Fokker-Planck models are valid only for scale-free density profiles and at
radii inside the \sbh\ sphere of influence.
Similar $N$-body studies \citep{BrockampBK2011, FiestasPBS2012}
typically assign a mass to the \sbh\ particle of
$\sim 10^{-3}-10^{-2}$ times the mass in  stars, similar to the observed ratio.
Here \citep[as in][]{KomossaMerritt2008} we adopt larger values for this ratio
in order to study in detail the region inside the influence sphere. 
We used two values for $\Mbh$: $0.1$ and $0.02$ times the mass in stars
(the latter set to unity). 
In order to simulate various evolutionary regimes (e.g. empty/full loss cone) we varied 
the radius $r_\mathrm{lc}$ at which stars are captured between $10^{-5}$ 
and $2\times 10^{-4}$ (in units of the Dehnen-model scale length), 
and we also varied the number of particles in the system:
$N=2.5\times 10^4, 10^5$ and $2.5\times 10^5$.
We stress that in no case would our models correspond to real galaxies
(the capture radius is too large and the number of stars too small), 
but once we understand the dependence of the evolution on 
these parameters, we can scale the results to real galaxies. 
We summarize the parameters of our models in Table~\ref{tab_models}.

\begin{table*}
\caption{Parameters of the $N$-body and Fokker-Planck models}  \label{tab_models}
$N$ is the number of particles, 
$r_\mathrm{lc}$ is the loss cone radius (distance to \sbh\ at which stars are captured),
$\trel$ is the central relaxation time defined in equation~(\ref{eq_Trel}),
$T_\mathrm{sim}$ is the duration of the simulation, 
$\ln\Lambda$ is the Coulomb logarithm,
$\rh$ is the influence radius,
$r_\mathrm{local(global)}$ are radii corresponding to energy $\calE_\mathrm{local(global)}$
of transition between empty and full-loss-cone regimes for the spherical problem
defined at the end of \S\ref{sec_boundary_cond_spher},
$N_\mathrm{capt}$ is the total number of particles captured by the end of integration 
(separately for spherical and axisymmetric case with axis ratio of 0.75, 
and for Fokker-Planck and $N$-body models).
\begin{tabular}{lrlllllllllll}
\hline
\hline
&&&&&&&&& \multicolumn{2}{c}{$N_\mathrm{capt}$, spherical} & \multicolumn{2}{c}{$N_\mathrm{capt}$, flattened} \\
Model & $N$ & $\Mbh$ & $r_\mathrm{lc}$ & $\trel$ & $T_\mathrm{sim}$ & $\ln\Lambda$ & 
$\rh$ & $r_\mathrm{local(global)}$ &
F-P & $N$-body & F-P & $N$-body \\
\hline
M1 & $10^5$ & 0.1 & $10^{-4}$ & 250 & 100 & 8 & 0.5 & 0.09\,(0.45) & 560 & 620 & 990 & 930 \\
M2 & $10^5$ & 0.1 & $10^{-5}$ & 250 & 100 & 8 & 0.5 & 0.03\,(0.09) & 190 & 240 & 300 & 280 \\
M3 & $2.5\times10^4$ & 0.1 & $10^{-5}$ & 75 & 50 & 6.6 & 0.5 & 0.017\,(0.045) & 45 & 45 & 55 & 60 \\
M4 & $2.5\times10^5$ & 0.1 & $2\times10^{-4}$ & 550 & 100 & 8.9 & 0.5 & 0.25\,(4.5) & 1160 & 1260 & 2550 & 2540 \\
M5 & $2.5\times10^5$ & 0.02 & $10^{-5}$ & 60 & 50 & 7.3 & 0.14 & 0.013\,(0.037) & 270 & 320 & 380 & 370 \\
\hline
\end{tabular}
 \end{table*}

For the flattened models we adopted a density axis ratio of $p=0.75$, 
which corresponds to $\calRsep \approx 0.29$ via equations 
(\ref{eq_epsilond}), (\ref{eq_CalHb}).
We did not vary this parameter since the foregoing analysis indicated a rather weak 
dependence of the flux on $\calRsep$ (e.g. equation~\ref{eq_capt_rate_axi}).

The flattened models were constructed with the \citet{Schwarzschild1979}
orbit superposition method,  in the implementation described in 
\citet{Vasiliev2013}, using $\sim10^5$ orbits.  
While there is a unique two-integral distribution function $f(\calE,L_z)$ that 
self-consistently reproduces a given $\rho(R,z)$ \citep{LyndenBell1962,HunterQian1993}, 
there are infinitely many three-integral distribution functions 
\citep[e.g.][]{DehnenGerhard1993}.
In order to construct models that were ``most similar'' to isotropic spherical
models, we chose the orbital weights in such a way as to minimize a global
measure of the ``velocity anisotropy'' 
$\beta \equiv 1-(\sigma_\theta^2+\sigma_\phi^2)/(2\sigma_r^2)$.
The resultant models are characterized by a non-trivial dependence of
$f$ on the third integral, since in a two-integral, $f(E,L_z)$ model,
velocities are forced to be isotropic in the meridional plane only, 
i.e. $\sigma_\theta^2=\sigma_r^2$, and $\beta\ne0$ in general.
Among the numerical checks that we carried out was to construct
spherical models using both orbital superposition, as well as 
Eddington's inversion formula; 
no noticeable difference was found in the $N$-body evolution.

The models were evolved using the direct-summation $N$-body code $\phi$GRAPEch 
\citep{HarfstGMM2008}, which uses algorithmic regularization 
\citep{MikkolaMerritt2006,MikkolaMerritt2008} 
to increase the speed and accuracy of particle
advancement near the \sbh, and includes an option for capturing particles that pass 
within a specified distance from the \sbh; the mass of captured stars is added
to $\Mbh$. 
Integrations were carried out both on GRAPE workstations and with the GPU-accelerated 
SAPPORO library \citep{GaburovHP2009}.
The accuracy parameter of the Hermite integrator was set to $\eta=0.01$ and 
the gravitational softening length $\epsilon$ was set to zero.
We used purely Newtonian gravity, as the influence of relativistic effects on the 
total capture rate (due mainly to stars with $a\sim\rh$) is likely to be
negligible (Appendix \ref{sec_appendix_relativity}).
The integration time $T_\mathrm{sim}$ in $N$-body units was chosen to be 
a substantial fraction of the central relaxation time $\trel$ (\ref{eq_Trel}), 
but not longer, in order to avoid significant changes in the density profile. 
In all integrations, the mass of captured stars was a small fraction of $\Mbh$, 
therefore we did not change $r_\mathrm{lc}$ as a function of time.
A star was considered to be captured if its angular momentum near periapsis 
passage was less than $L_\mathrm{lc} = 
r_\mathrm{lc}\sqrt{2(G\Mbh/r_\mathrm{lc}-\calE)} \approx \sqrt{2G\Mbh r_\mathrm{lc}}$;
we used the latter approximation which is valid for highly eccentric orbits.
Angular momentum is preferable to periapsis radius as a condition for capture since
$L$ can be computed far from the \sbh\ particle where ambiguities due to GR
are negligible.

Corresponding Fokker-Planck models were constructed in the following way. 
We used Eddington's formula to obtain the isotropized distribution function $f(\calE)$
for the given density profile. 
For each value of energy the diffusion in angular momentum (for the spherical case) 
or in the $\calH-\calR_z$ space (in the flattened case) was considered using 
the analytical expressions for the time-dependent one-dimensional solution in terms of 
Bessel functions \citep{MilosMerritt2003}, or the steady-state expressions 
(\ref{eq_f_global_spher}-b). 
For the flattened system we used the equivalent 1d prescription from 
\S~\ref{sec_FP_solution}, with $\alpha_\mathrm{axi}$ and $\calReff$ given 
by equations (\ref{eq_alpha_axi}), (\ref{eq_Reff_axi}).
We also accounted for the draining of chaotic orbits and the loss wedge using the 
expressions (\ref{eq_Fchaotic}), (\ref{eq_Tdrainch}) for the draining rate of an initially 
full loss region (i.e. isotropic distribution) and the approximate combination of draining 
and relaxation capture rates described at the end of \S~\ref{sec_chaos}; 
the fraction of chaotic orbits $\calRch$ was set equal to $\calRsep$. 
Thus our Fokker-Planck models represent the same starting conditions as the $N$-body integrations.

We used a number of criteria for comparing the results from the Fokker-Planck 
and $N$-body models.
Below we present comparison for some of these criteria evaluated for model M1 
(spherical and axisymmetric cases), although the results were similar for other models.

\begin{figure}
$$\includegraphics[angle=-90]{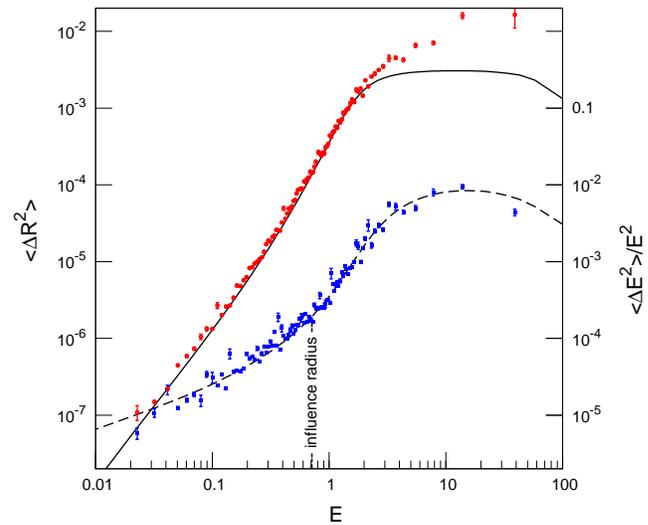}$$
\caption{
Comparison of theoretical diffusion coefficients with results from the 
$N$-body simulations (spherical Dehnen model with $\Mbh=0.1$, $N=10^5$). 
Theoretical coefficients are shown by the solid ($\calR$) and dashed ($\calE$) curves.
The diffusion coefficient in energy has been shifted downward by two decades to avoid overlap. 
} \label{fig_dif_coefs}
\end{figure}

\begin{figure*}
$$\includegraphics[angle=-90]{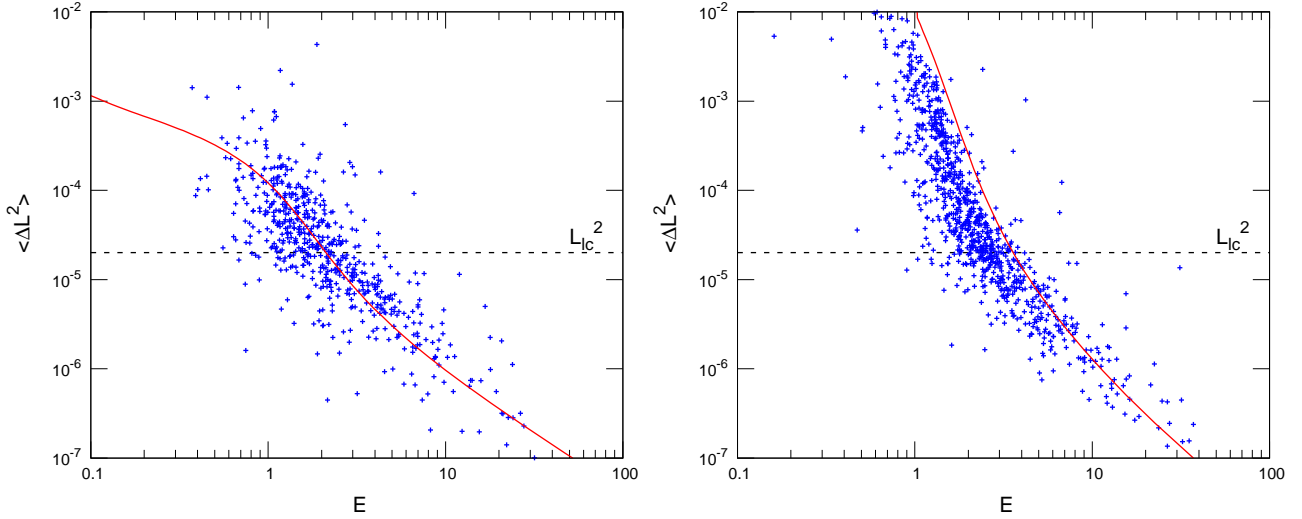}$$
\caption{
Change in squared angular momentum during the final orbit before capture, 
for spherical (left) and axisymmetric (right) models with 
$\Mbh=0.1$, $N=10^5$, $r_\mathrm{lc}=10^{-4}$. 
Solid lines are predictions from loss cone theory: 
equation (\ref{eq_Delta_R_sph}) for the spherical case, 
equation (\ref{eq_Delta_R_axi}) for the axisymmetric case;
the latter should be regarded as an 
upper limit for the reasons discussed in the text.
Horizontal dashed line is the loss cone boundary, $L^2=2G\Mbh r_\mathrm{lc}$
}  \label{fig_delta_l}
\end{figure*}

The first indicator is the rate of relaxation in energy and angular momentum.
In the simulations, we sorted all particles in initial $\calE$ and $\calR$ and 
divided them into 100 bins, 
then computed $(\Delta\calR)^2$ and $(\Delta\calE)^2/\calE^2$ for each particle
and averaged these values within each bin. For a diffusive process, these quantities 
should grow linearly with time and so we fitted the time dependence with a straight line
and took the slope of this fit as the measured value of $\langle (\Delta\calE)^2\rangle,\langle(\Delta \calR)^2\rangle$.
 Theoretical diffusion coefficients were computed by averaging the local 
coefficients over the volume of phase space accessible at a given energy
(equation \ref{eq_dif_coefs_avg_E}).
The comparison between these theoretical coefficients and the values measured from 
the simulations is shown in Figure~\ref{fig_dif_coefs} for the (spherical) model M1.
The Coulomb logarithm is the only adjustable parameter in this comparison, 
and the best agreement was obtained setting it roughly equal 
to the logarithm of the number of particles inside the influence radius,
$\ln\Lambda \approx \ln (0.3 \Mbh/m_\star)$ 
\citep[e.g.][equation 5.35]{DEGN}.
Agreement was found to be good beyond and just inside the \sbh\ sphere of influence,
$\calE\approx 1$,
while at smaller radii (larger binding energies) relaxation in angular momentum 
was found to be faster than predicted, which may be an indication of resonant relaxation 
\citep{RauchTremaine1996}, 
although the diffusion rate was not as high as measured by \citet{EilonKA2009}.
For a power-law cusp with $\gamma=3/2$, both $\langle(\Delta\calR)^2\rangle$ 
and $\langle(\Delta\calE)^2\rangle/\calE^2$ 
should tend to constant limits for $\calE\to \infty$. 
Since the number of stars in the simulations is finite and there is a maximum value of 
the binding energy, $\calE \approx 100$, for an $N=10^5$ particle system, we introduced 
an upper energy cutoff in the distribution function in computing the theoretical coefficients
\citep[cf.][Appendix D]{BarorKA2013}, which results in the decline of the coefficients at 
large $\calE$. 
We did not attempt to study resonant relaxation in more detail since it is not well 
described by our Fokker-Planck formalism and more sophisticated statistical models may 
be needed.
In any case, the enhancement in the capture rate due to resonant relaxation is 
expected to be small due to the small number of particles at high binding energies \citep{HopmanAlexander2006}.

Next we compare the properties of captured particles and the population of the 
loss cone with the predictions of the Fokker-Planck models, in both spherical
and axisymmetric models.
For every captured star, we recorded the energy and angular momentum at the 
moment of capture, then looked back to find their changes since the previous 
periapsis passage. 
Figure~\ref{fig_delta_l} plots changes in squared angular momentum 
during the final orbit versus particle energy and compares it with the expected 
(average) change due to diffusion.
In the case of spherical models (Figure~\ref{fig_delta_l}a),
those changes were predicted in terms of $q(\calE)$ 
(\ref{eq_q_spher}) as follows:
\begin{subequations}
\begin{eqnarray}  \label{eq_Delta_R_sph}
\Delta\calR &=& \sqrt{\Dif_{RR}\,\trad} = 
  \sqrt{\left(\frac{\Dif_{RR}}{\calR}\right) (\calRcapt+\Delta\calR)\,\trad} = \nonumber\\
&=&  \sqrt{q\calRcapt(\calRcapt+\Delta\calR)} =
  q\calRcapt \frac{1+\sqrt{1+4q^{-1}}}2,  \nonumber\\ \\
(\Delta L)^2 &=& \Lcirc^2\Delta\calR = 2G\Mbh r_\mathrm{lc}\, q\frac{1+\sqrt{1+4q^{-1}}}2
  \label{eq_delta_L2_last} .
\end{eqnarray}
\end{subequations}
In spite of the substantial scatter, the measured angular momentum 
changes are well described by this approximation. 

For the axisymmetric case (Figure~\ref{fig_delta_l}b)
the angular momentum changes during the final orbit are 
higher due to torques from the flattened potential.
We can estimate $\langle\Delta\calR\rangle$ by approximating the time 
evolution of $\calR$ (\ref{eq_calR_of_t}) near the minimum as a parabola, 
taking $\calRmax=\calRsep, \calRmin=0$ and evaluating the difference
\begin{equation}
\Delta\calR = (\calRmax-\calRmin)\frac{\pi^2}{\tosc^2} 
\left[(T_\mathrm{lc}+\trad)^2-T_\mathrm{lc}^2\right] \;,
\end{equation}
where $T_\mathrm{lc}\equiv (\tosc/\pi)\sqrt{\calRcapt/\calRmax}$ 
denotes the elapsed time after entering the loss cone until reaching 
the minimum $\calR$. This estimate gives an upper limit to $\Delta\calR$, 
expressed in terms of coefficient $q_\mathrm{axi}$ (\ref{eq_q_axi}):
\begin{equation}  \label{eq_Delta_R_axi}
\Delta\calR = \calRcapt\times \pi q_\mathrm{axi}(2q_\mathrm{axi}+\pi) ,
\end{equation}
which is plotted as a solid line in Figure~\ref{fig_delta_l}b; 
the measured values of $\Delta\calR$ indeed lie below this upper limit 
but are higher than in the spherical case.

\begin{figure}
$$\includegraphics[angle=-90]{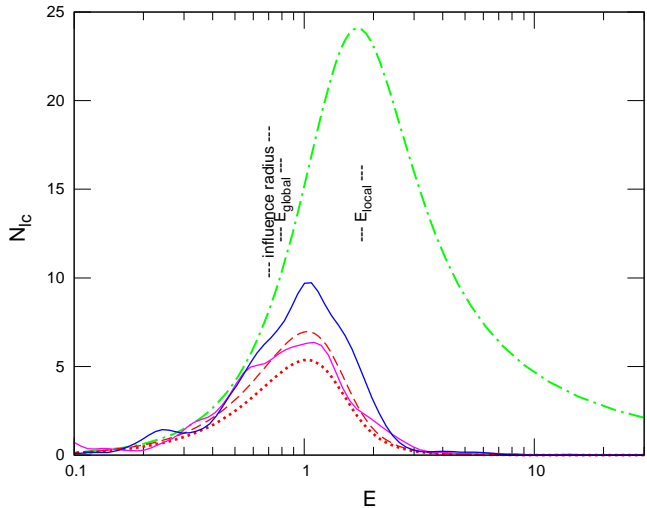}$$
\caption{
Instantaneous number of stars in the loss cone as a function of energy.
Solid curves are derived from the $N$-body simulations:
top (blue) -- axisymmetric; bottom (magenta) -- spherical (same model as 
in the previous figure).
Green dot-dashed curve is for a full loss cone (equation~\ref{eq_losscone_num_full} 
with $\overline \calN(\calE)$ equal to its initial value).
Dotted red curve is the ``real'', stationary loss cone population
(equation ~\ref{eq_losscone_num_real})
and the dashed red curve is the time-dependent population,
both for the spherical problem. 
} \label{fig_losscone_population}
\end{figure}

The population of the loss cone in the $N$-body simulations was 
computed as the instantaneous 
number of stars having angular momenta less than $L_\mathrm{lc}$. 
The corresponding quantity in the Fokker-Planck models is 
\begin{equation}  \label{eq_losscone_num}
N_\mathrm{<lc}(\calE)\,d\calE = \int_0^{\calRcapt(\calE)} 
  \calN(\calE,\calR)\, d\calR \,d\calE .
\end{equation}
We distinguish between $\Nfull$ -- the number of stars if their 
distribution in squared angular momentum is uniform (i.e. if the loss cone is 
full and the value of distribution function $\calN_\mathrm{lc}$ for $\calR<\calRcapt$ 
is the same as the isotropic value $\overline \calN$), 
and $\Nreal$ -- the number of stars if $\calN(\calE,\calR)$ 
is taken from the true solution or its quasi-steady-state approximation 
(equation \ref{eq_f_global_spher}).
Neglecting the variation of orbital period with $\calR$ at given $\calE$, 
the former quantity becomes 
\begin{subequations}
\begin{equation}  \label{eq_losscone_num_full}
\Nfull(\calE) \approx \calRcapt \overline \calN(\calE),
\end{equation}
and the latter (steady-state) is 
\begin{equation}  \label{eq_losscone_num_real}
\Nreal=\Nfull\,\frac{\exp(-\alpha)-1+\alpha}{\alpha-\ln\calRcapt-1} ,
\end{equation}
\end{subequations}
where $\alpha(q(\calE))$ is given by (\ref{eq_alpha_q}).
Figure~\ref{fig_losscone_population} shows the distribution of loss cone stars in 
energy in the simulations; overplotted are curves corresponding to full and real 
loss cone in time-dependent and steady-state spherical Fokker-Planck solutions.
We did not derive analogous expressions for the axisymmetric case, 
but the results from the simulations indicate that in the latter case the number of 
stars in the loss cone is somewhat higher than in the spherical system for 
energies $\calE \ge \calE_\mathrm{global}$, i.e. above the transition from 
full to empty-loss-cone regimes in the spherical system.

\begin{figure}
$$\includegraphics[angle=-90]{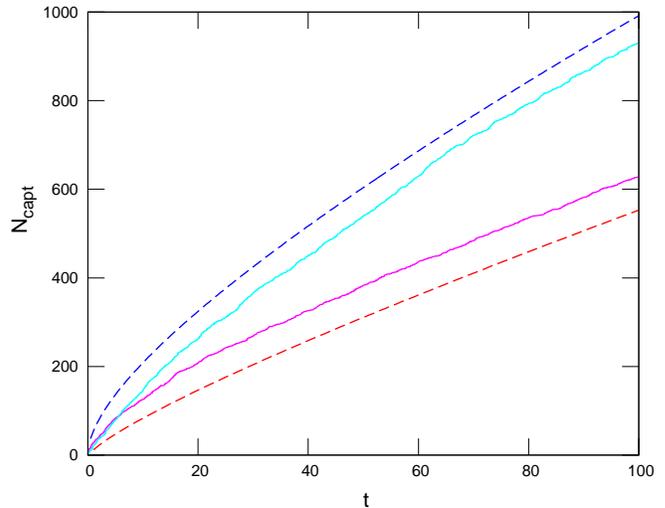}$$
\caption{
Cumulative number of captured stars in the simulations (run M1) as a function of time (points), 
compared with predictions from the time-dependent Fokker-Planck models (dashed lines).
Top (blue) curves: axisymmetric model; bottom (red/magenta) curves: spherical model.
} \label{fig_capt_total}
\end{figure}

\begin{figure}
$$\includegraphics[angle=-90]{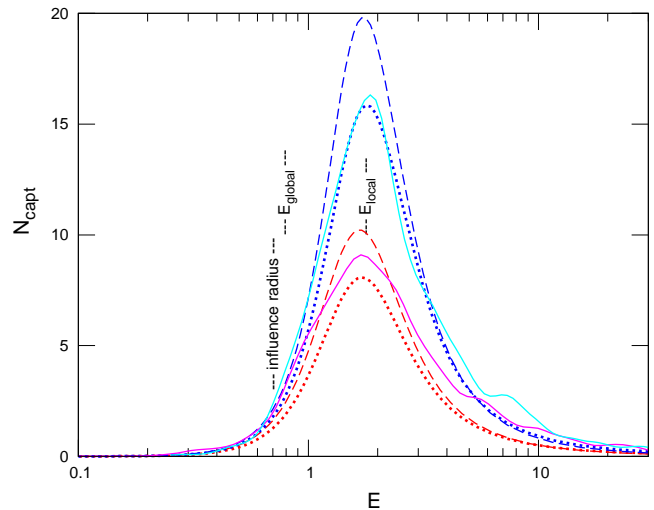}$$
\caption{
Distribution of captured stars in energy (plotted is the number of stars 
captured per unit time).
Solid curves are from the $N$-body integrations: 
top (blue): axisymmetric; bottom (magenta): spherical models, in both cases with 
$\Mbh=0.1$, $N=10^5$, $r_\mathrm{lc}=10^{-4}$ (model M1). 
Overplotted are predictions from the Fokker-Planck models: 
dotted: stationary flux (equation~(\ref{eq_capt_rate_spher}) for the spherical case (red), 
equation~(\ref{eq_capt_rate_axi}) for the axisymmetric case (blue) with the effective 
capture boundary $\calReff$ given by equation~(\ref{eq_Reff_axi})); 
dashed: time-dependent solution of the 1d spherical problem for the same two cases.
}  \label{fig_capt_distrib}
\end{figure}

Finally, we consider the capture rate, or, rather, the cumulative number of 
stars captured since the beginning of the simulation as a function of time.
The stationary solution of the 1d Fokker-Planck equation underestimates 
the capture rate at early times when the phase space density near the loss cone 
has not yet reached its steady-state value, so we adopted the time-dependent 
solution as a basis for comparison.
Figure~\ref{fig_capt_total} shows that the capture rate decreases with time, 
i.e. the cumulative number of captured stars grows more slowly than linearly,
as expected.
Figure~\ref{fig_capt_distrib} shows the distribution of captured particles in energy,
which matches the Fokker-Planck solution very well, apart from an excess of 
captured particles at high binding energies in the simulations, which is a
consequence of the higher rate of diffusion in angular momentum discussed above 
(Figure~\ref{fig_dif_coefs}).
Overall, the capture rate for model M1 in the axisymmetric case is $\sim 50\%$ 
higher than for the spherical model with the same density profile, 
confirming the predictions of the Fokker-Planck study.
The same was found to be true in the other models of Table~\ref{tab_models}: 
flattening was never found to make more than a factor of two difference. 
Moreover, models for which the transition to the full-loss-cone regime in the 
spherical problem occurs well within the radius of influence (M2, M3, M5) showed 
less difference than the model M4 which is mainly in the empty-loss-cone regime;
these results are in line with theoretical predictions presented at the end of \S~\ref{sec_FP_solution}.

\begin{figure}
$$\includegraphics[angle=-90]{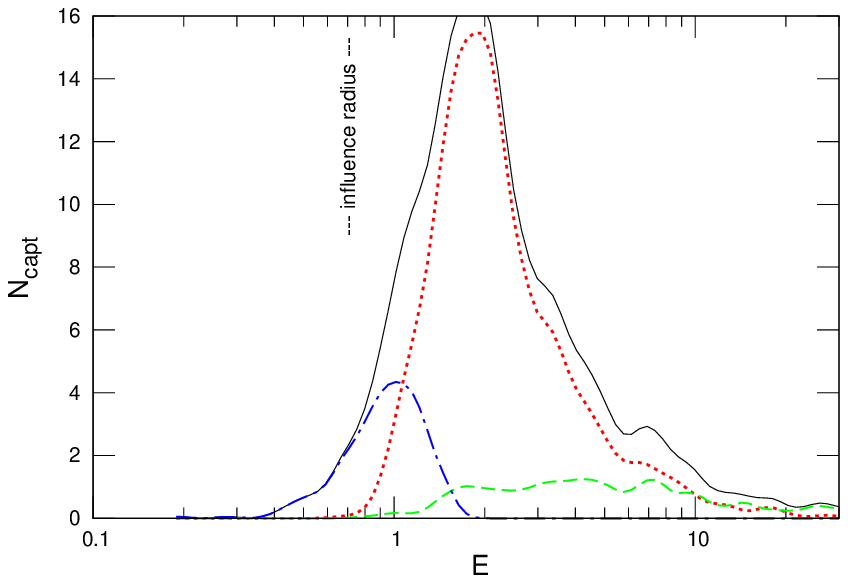}$$
$$\includegraphics[angle=-90]{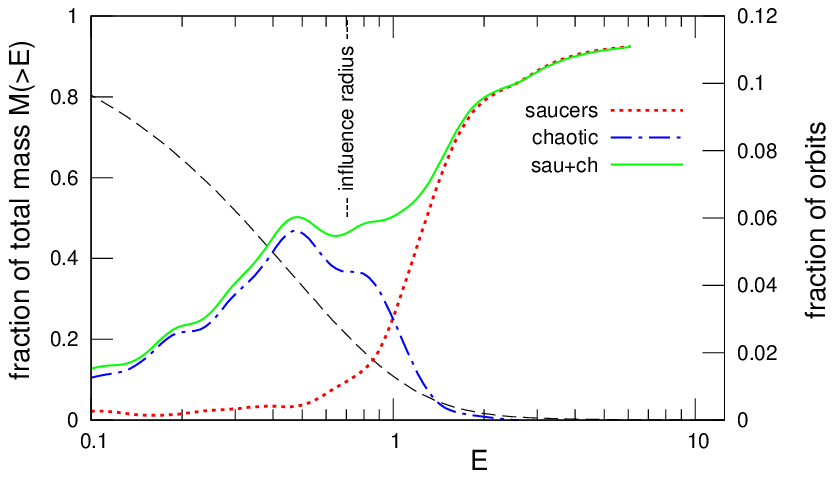}$$
\caption{
Top panel: Distribution of captured stars (per unit of time) by orbital type
for axisymmetric model M1.
Dotted red: saucers; dashed green: tubes; dot-dashed blue: chaotic;
solid line: total. 
Bottom panel: Fraction of different types of orbits in the axisymmetric Schwarzschild model M1, 
as a function of energy.
Dotted red: saucers; dot-dashed blue: chaotic (numbers are on the right-hand ordinate);
solid green: sum of these two; 
dashed black: cumulative mass (fraction of stars with energy $>\calE$,
numbers on the left ordinate).
} \label{fig_orbit_type}
\end{figure}

We also recorded the orbital parameters of captured stars at their final
apoapsis passage and followed the orbits in the smooth potential used 
to construct the flattened models. Figure~\ref{fig_orbit_type}, top panel, shows that 
most of these stars found their way into the \sbh\ while being on saucer orbits, 
a result also predicted by the axisymmetric Fokker-Planck models. 
The bottom panel of this figure shows that around and beyond the radius of influence, 
chaotic orbits play a similar role to saucers, however, in this particular model their 
contribution to the total capture rate is small.

As a final remark, we tested that the $N$-body models were in dynamical 
equilibrium by examining the evolution of Lagrangian radii of shells containing 
given fractions of the total mass (5\%, 10\%, etc.), and also the axis ratios of the models. 
In integrations with captures disabled, these did not change apart from small fluctuations.
When captures were enabled, 
Lagrangian radii expanded slightly with respect to time 
(corresponding to  energy input from the \sbh), 
while the axis ratios did not change appreciably.
The \sbh\ particle did not remain precisely at the model center but rather experienced 
Brownian motion \citep{MerrittBerczikLaun2007}; 
however the amplitude was at least an order of magnitude smaller than 
the influence radius. \citet{BrockampBK2011} found no substantial 
differences in capture rate between simulations with fixed and wandering \sbhs.

\section{Estimates for real galaxies}  \label{sec_estimates}

\begin{figure*}
$$\includegraphics[angle=-90]{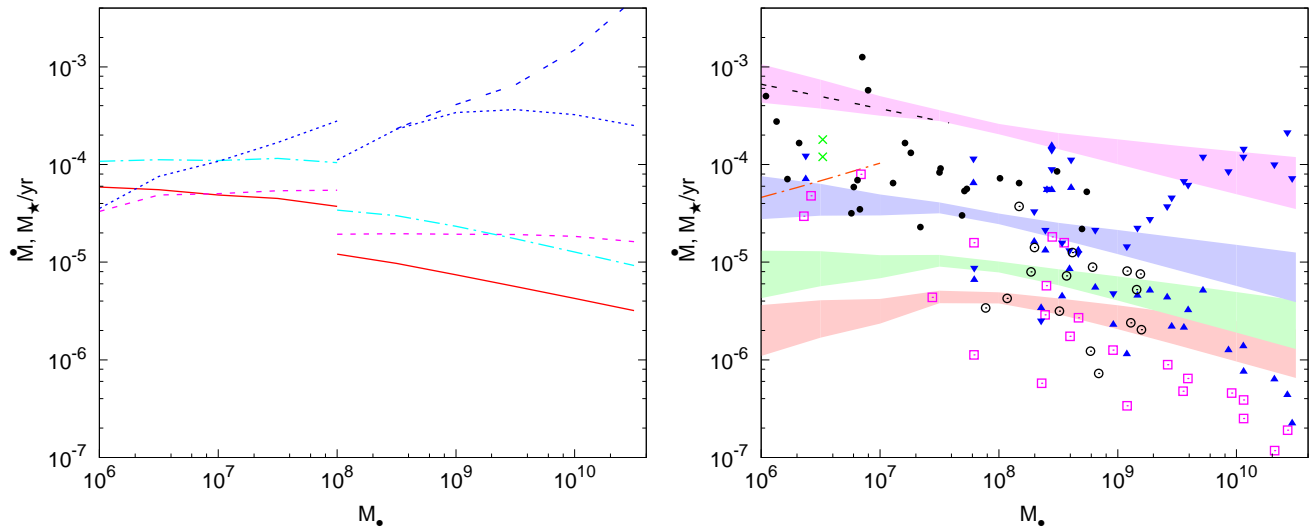}$$
\caption{
Estimates of the capture rate $\dot M$ as a function of \sbh\ mass $\Mbh$ and of the  parameters defining the galaxy. 
{\it Left panel}: Predictions from Fokker-Planck models for galaxies with inner density cusp slope 
$\gamma=1.5$ for $\Mbh\le 10^8\,M_\odot$ and $\gamma=1$ for $\Mbh\ge 10^8\,M_\odot$ 
(the discontinuity corresponds to $\sim 3$ times higher rates for more concentrated galaxies). 
Galaxy models are scaled to match the $\Mbh-\sigma$ relation (\ref{eq_MbhSigma}) with $\alpha=8,\beta=4.5$.
Solid red curves and dashed purple lines: stationary and time-dependent flux in the spherical case;
dot-dashed and dotted blue lines: same for axisymmetric case with $\calRsep=0.1$;
double-dashed blue line: flux in the axisymmetric case with contribution from loss region draining.
On average, stationary capture rates differ by a factor of 2-3 between spherical and axisymmetric cases; the
time-dependent solution generally yields higher rates since it starts from an initial condition with 
strong gradients in the distribution function near the loss region, which may not be physically motivated 
and hence provides an upper limit to possible rates in real galaxies.
{\it Right panel}: Comparison of stationary capture rates in the spherical case for different density cusp slopes; 
shaded regions correspond to uncertainty due to the range of parameters
$\{\alpha,\beta\}$ in the $\Mbh-\sigma$ relation (equation \ref{eq_MbhSigma}, Table~\ref{tab_Msigma}). 
From top to bottom, $\gamma=2,1.5,1,0.5$. 
Points are taken from previous studies, as follows.
Purple open boxes: \citet{SyerUlmer1999}; 
blue upward and downward triangles: \citet{MagorrianTremaine1999} for the spherical and axisymmetric cases;
black open and filled circles: \citet{WangMerritt2004} for cored and cuspy galaxies; 
black double-dashed line: the analytic estimate from \citet{WangMerritt2004}
for a singular isothermal sphere ($\gamma=2$);
red dot-dashed line: \citet{BrockampBK2011} using their extrapolation from $N$-body simulations 
of $n=4$ S\'ersic galaxy models;
green crosses: \citet{FiestasPBS2012} for $N$-body evolution of King models scaled to 
Milky Way nuclear star cluster with a $\gamma=1.75$ cusp (spherical and flattened case).
} \label{fig_rates}
\end{figure*}

Having verified that the Fokker-Planck models agree well with the $N$-body simulations 
in the range of parameters feasible for the latter, we now extrapolate the Fokker-Planck 
theory to parameters characteristic of real galaxies.
To that end, we consider a family of  models represented by 
\citet{Dehnen1993} density profiles having inner cusp slopes $0.5\le \gamma \le 2$, 
and take the \sbh\ mass to be $10^{-3}$ of the total galaxy mass
\citep{MerrittFerrarese2001,MarconiHunt2003}.
The corresponding influence radii lie well inside the break radius separating
the inner $\rho\sim r^{-\gamma}$ cusp from the outer $\rho\sim r^{-4}$ profile, so that only 
the density normalization  at the radius of influence (say) matters.
That density is, in principle, an independent parameter but we fix it 
via the requirement that the galaxy satisfy the so-called 
$\Mbh-\sigma$ relation linking $\Mbh$ to the velocity dispersion observed 
near the center of the galaxy. 
The $\Mbh-\sigma$ relation is typically written in the form
\begin{equation}  \label{eq_MbhSigma}
\log \Mbh = \alpha + \beta \log(\sigma/200\,\mathrm{km\ s}^{-1}) ,
\end{equation}
with parameters $\alpha \approx 8$, $4\lesssim \beta \lesssim 5$, 
depending on the details of sample selection (Table~\ref{tab_Msigma}).
\begin{table}
\caption{Parameters of the $\Mbh-\sigma$ relation}  \label{tab_Msigma}
\begin{tabular}{lll}
\hline
$\alpha$ & $\beta$ & Reference \\
\hline
8.49 & 4.00 & \citet{Merritt1999} \\
8.08 & 3.75 & \citet{Gebhardt2000} \\
8.22 & 4.86 & \citet{FerrareseMerritt2000} \\
8.13 & 4.02 & \citet{Tremaine2002} \\
8.12 & 4.24 & \citet{Gultekin2009} \\
8.13 & 5.13 & \citet{GrahamOAC2011} \\
8.29 & 5.12 & \citet{McConnell2011} \\
\hline
\end{tabular}
\end{table}

We scale the velocity unit of the Dehnen model by identifying
the line-of-sight velocity dispersion, $\sigma_p(R)$, 
at the influence radius with the quantity $\sigma$ in equation~(\ref{eq_MbhSigma}); 
this is reasonable given that $\sigma_p$ is a weak function of projected
radius $R$ for $\rh\lesssim R \lesssim R_e$, with $R_e$ the half-light radius.
(In the case $\gamma=2$ the central $\sigma_p$ in the model lacking a \sbh\
was used.)
Our models are thus defined by the two parameters $\{\Mbh, \gamma\}$. 
Setting $\alpha=8,\beta=4.5$, we have 
$\rh=\{65,45,25,11\}\times (\Mbh/10^8\,M_\odot)^{0.56}$~pc for 
$\gamma=\{1/2,1,3/2,2\}$. 

The radius $r_\mathrm{lc}$ that defines the loss sphere around the \sbh\
is the larger of the radius of tidal disruption, $r_\mathrm{tid}$, and
the (Newtonian) periapsis of an orbit
that just continues inside the event horizon.
For the eccentric orbits that dominate the flux into the \sbh, 
the latter quantity is $\sim 8\rg=8 G\Mbh/c^2$ for a 
Schwarzschild (nonrotating) \sbh\ \citep{Will2012};
this is the radius of periapsis of a Keplerian orbit having the critical angular momentum. 
A star is tidally disrupted if the periapsis radius is less than 
\begin{equation}
\rtid \approx \rg\times 2.2\eta^{2/3} \left(\frac{\Mbh}{10^8\,M_\odot}\right)^{-2/3} 
  \left(\frac{m_\star}{M_\odot}\right)^{-1/3} \frac{r_\star}{r_\odot} 
\end{equation}
\citep[e.g.][equation 6.3]{DEGN}.
Here $\eta$ depends on the stellar equation of state and is $\sim0.84$ for a solar-type 
main-sequence star. 
These disruption events, as opposed to direct captures, may be observed as optical and 
x-ray flares in otherwise quiescent galactic nuclei \citep{StrubbeQuataert2009}.
From the condition $\rtid>8\rg$ we find that solar-type stars on eccentric orbits
are disrupted (not swallowed) if $\Mbh\lesssim 1.2\times 10^7\,M_\odot$;
disruption can occur for any $\Mbh\lesssim 10^8M_\odot$ if the star is on 
a less eccentric orbit, or for Kerr \sbhs\ even more massive than $10^8\,M_\odot$ 
\citep{Kesden2012}.
Red giants or AGB stars can also be disrupted (or at least tidally limited) 
by \sbhs\ more massive than $10^8\,M_\odot$ \citep{SyerUlmer1999, MacleodGR2012}.
In what follows, we compute the total number of events associated with a given
$r_\mathrm{lc}$; 
the ratio of number of tidal disruption flares to the total number of capture events 
is well studied in the literature and we do not consider it separately here.

We used Fokker-Planck models to evaluate steady-state and time-dependent capture rates $\dot M$ 
for galaxies after $10^{10}$ years starting from an initially isotropic distribution function,
in both the spherical and axisymmetric geometries (using our one-dimensional approximation of 
section~\ref{sec_FP_solution}), for $\calRsep=0.1$.
The latter value is meant to represent a ``typical,'' moderately-flattened system; 
the results do not strongly depend on $\calRsep$.
In the time-dependent calculations, the initial conditions consisted of the
isotropic models with loss-cone orbits removed; as a result, these initial
models are characterized by strong gradients of $f$ with respect to $L$ near the loss cone.
Overall, the calculation of total capture rate is done in the same way as in the previous section.

The left panel of Figure~\ref{fig_rates} shows results for two series of models:
models with a steep ($\gamma=1.5$) central cusp and $\Mbh\le 10^8\,M_\odot$;
and models with shallow ($\gamma=1$) cores and $\Mbh\ge 10^8\,M_\odot$.
Over the entire range of $\Mbh$, the steady-state capture rates differ by only a factor of 
$2-3$ between spherical and axisymmetric geometries, consistent with the discussion 
near the end of \S\ref{sec_FP_solution}. 
This result holds for any galactic model and depends only weakly on $\calRsep$.
The time-dependent rates are generally higher than in the steady state, 
due to the strong gradients in the initial conditions.
Especially for massive 
galaxies with long relaxation times, the approach to a steady state is slow and 
the flux at early stages is much higher than in equilibrium.

For the most massive \sbhs\ ($\Mbh \gtrsim 10^9\,M_\odot$) the draining time of the 
loss region  becomes comparable to the Hubble time and the capture rate is dominated by 
draining of chaotic orbits (double-dashed line),
reaching values up to $10^{-3}\,M_\odot$ yr$^{-1}$ 
at the upper end of the $\Mbh$ range.
However, this should be regarded as a strong upper limit since we do not know 
the initial state: for instance, if the \sbh\ formed as the result of a merger 
of a binary \sbh, it is very likely that the low angular momentum region of phase space 
will have been depleted in the course of the binary's evolution, and the capture rates could 
initially be much {\it lower} than the steady-state values \citep{MilosMerritt2003,MerrittWang2005}.

The dependence of $\dot M$ on $\Mbh$ for massive \sbhs\
can be simply estimated as follows \citep[e.g.][]{SyerUlmer1999}. 
The nuclei of massive galaxies are in the empty-loss-cone regime, so 
the flux per unit energy is roughly $\calF(\calE)\,d\calE \approx 
\calN(\calE)\,d\calE / \left[T_\mathrm{rel}(\calE)\,\ln \calRcapt^{-1}(\calE)\right]$. 
The capture rate peaks at $r\approx \rh$ \cite[][section 6.1.4.1]{DEGN}
so the total flux is estimated as 
$\dot M \sim \calF(\calE_\mathrm{infl})\calE_\mathrm{infl} \sim 
\Mbh/(T_\mathrm{rel} \ln \calRcapt^{-1}) \sim (m_\star/\Mbh)\,\sigma^3/G$. 
Assuming the $\Mbh-\sigma$ relation, we obtain $\dot N \propto \Mbh^{3/\beta-1}$, 
and even the normalization constant evaluates to a reasonable $\sim 10^{-5}\,M_\odot$ yr$^{-1}$ 
for $\Mbh=10^8\,M_\odot$, despite the crudeness of the estimate.

The right panel of Figure~\ref{fig_rates} shows the uncertainties in the capture rate
associated with the  parameters $\alpha$ and $\beta$ in the $\Mbh-\sigma$ relation 
(\ref{eq_MbhSigma}) and the slope of the density cusp $\gamma$. 
Plotted are stationary capture rates for the 
spherical case; other values scale roughly in the same proportion. 
The capture rates evaluated for a selection of individual galaxies from several previous 
studies are also plotted for comparison. 
It is clear that the scatter in the derived values is fairly large, about two orders of magnitude, 
although a general trend of decreasing  rate with increasing $\Mbh$ is clear.
Comparing the inverted triangles in Figure~\ref{fig_rates}b with the double-dashed
curve in Figure~\ref{fig_rates}a, we see that our estimates for the capture rate due
to draining of chaotic orbits are substantially higher than those of \citet{MagorrianTremaine1999}.
As argued above, this is most likely an overestimate 
resulting from simplistic initial conditions.
It is also due partly  to our selection of a different relation between the capture 
rate and the isotropized distribution function; had we
used their equation (\ref{eq_fluxMT}) instead of our equation (\ref{eq_capt_rate_axi_approx}),
the steady-state flux in the axisymmetric case would be factor of a few lower, although it should not 
affect the draining rate of centrophilic orbits which starts to dominate the capture rate at 
$\Mbh\gtrsim 10^9\,M_\odot$. 

Overall, it is fair to say that our estimates predict capture rates in the range 
$10^{-5}-10^{-4}\,M_\odot$ yr$^{-1}$ for less massive galaxies, and 
a few$\times 10^{-6}-10^{-5}\,M_\odot$ yr$^{-1}$ for giant galaxies with \sbh\ 
masses in excess of $10^8\,M_\odot$. 
This is roughly consistent with the observationally derived estimates of rates of 
tidal disruption flares \citep{Donley2002, Gezari2008, VelzenFarrar2012}.
(Axisymmetric) nuclear flattening may increase these numbers by a factor of few, and triaxiality 
may have a more 
dramatic effect on the consumption rate of the most massive \sbhs\ 
provided there are enough stars on centrophilic orbits. 

One potentially important feature of axisymmetric (and triaxial) systems is that most stars 
are consumed in the full-loss-cone regime of boundary conditions (in spite of the fact
that the angular momentum need not be isotropic). This means that stars approach the 
\sbh\ with a wide distribution in periapsis radii, as opposed to ``barely touching'' the 
disruption sphere in the empty-loss-cone regime. As a consequence, many stars will be strongly 
tidally distorted before disruption, which may result in a distinct observational signature 
\citep{StrubbeQuataert2009},
although more recent studies show that the difference may not be so pronounced \citep{StoneSL2013}.

\section{Conclusions}  \label{sec_conclusions}

We have considered collisional (gravitational-encounter-drive) relaxation processes 
near supermassive black holes (\sbhs) in spherical and axisymmetric models of
galactic nuclei.
We derived a Fokker-Planck formalism and compared its predictions with direct
$N$-body simulations of capture.

Inside the \sbh\ radius of influence, the unperturbed motion of stars is regular, admitting 
three integrals of motion (energy, $z$-component of the angular momentum $L_z$,
and secular Hamiltonian $\calH$).
There are two families of orbits, tubes and saucers; the latter exhibit large angular
momentum variations and stars on saucer orbits approach much more closely to the
\sbh\ than would be expected based on their average angular momentum.
Regularity of the motion allowed us to write down the orbit-averaged Fokker-Planck equation 
and to calculate the diffusion coefficients based on the standard formalism. 
We discussed the appropriate boundary conditions for  capture by the \sbh, 
and numerically solved the two-dimensional ($\calH-L_z$) Fokker-Planck equation. 
We showed that its solution can be well approximated by an equivalent 
one-dimensional solution for diffusion in angular momentum, given appropriate 
boundary conditions. 
An important difference with the spherical case is that the boundary condition 
at the loss region typically corresponds to the full-loss-cone regime,
in the sense that a change in angular momentum per one orbital period is larger 
than the size of the loss cone. Nevertheless, the global shape of the solution, 
and, consequently, the capture rate of stars, is determined by the diffusion 
coefficient (inversely proportional to the relaxation time), and depends only weakly 
on the effective size of the loss region.
This treatment was not entirely self-consistent, as it did not account for 
resonant relaxation or for the breakdown of the orbit-averaged approximation 
near or beyond the radius of influence, but it nevertheless suggests a conclusion 
which turns out to be robust:
compared with the spherical case having the same \textit{real} 
(not \textit{effective}) capture boundary, the flux into the loss cone is higher
in the axisymmetric case, but not by a large factor, and only in the regime where
the loss cone would be empty in the spherical system. 
In the axisymmetric case, most of the stars find their way into the \sbh\ while on saucer orbits.

We also carried out a number of $N$-body integrations which were found to
agree remarkably well with the corresponding Fokker-Planck models.
The agreement was not limited to the number of captured stars; other quantities
like the distribution of energy of the captured stars, the loss cone population, 
the diffusion coefficients, and the change in angular momentum during one radial period 
prior to capture were also found to be well reproduced.
We note, however, that the angular momentum diffusion is larger in $N$-body models 
for stars with high binding energies (close to the \sbh), which may be an indication 
of resonant relaxation not accounted for in the Fokker-Planck models;
nevertheless, its influence on the capture rate is likely to be small.

We applied our Fokker-Planck formalism to realistic galaxies with \sbh\ masses 
$\Mbh=(10^6-10^{10})M_\odot$. 
We found that stationary capture rates are in the range 
$(10^{-4}-10^{-6})M_\odot$ yr$^{-1}$ in spherical galaxies and a 
factor $2-3$ higher in flattened systems, with an overall trend of 
decreasing event rate with increasing $\Mbh$.
Time-dependent solutions were found to give generally 
higher estimated capture rates; however, that result is likely to depend strongly 
on the assumed initial conditions \citep{MerrittWang2005}. 
In particular, for massive ($\Mbh \gtrsim 10^9\,M_\odot$) 
\sbhs\ in axisymmetric  nuclei, the draining time of chaotic orbits just outside the radius 
of influence can be comparable to the Hubble time; if such orbits were not depleted by
(for instance) the binary \sbh\ that preceded the single \sbh, capture rates might reach 
$10^{-3}\,M_\odot$yr$^{-1}$. 
Such high rates may be more relevant to triaxial galaxies in which the number of centrophilic 
orbits is likely to be large enough to maintain a full-loss-cone capture rate for a Hubble time.

\bigskip
The work was supported by the National Science Foundation via grant no.\ AST 1211602 
and by the National Aeronautics and Space Administration via grant no.\ NNX10AF84G.
EV acknowledges the hospitality of the Aspen Center for Physics.
Codes for computing capture rates in the spherical and axisymmetric geometries can be downloaded at http:$\slash\slash$td.lpi.ru$\slash$$\sim$eugvas$\slash$losscone$\slash$.

\appendix

\section{Effect of relativity on the unperturbed modtion}   \label{sec_appendix_relativity}

We discuss briefly the character of the  motion in axisymmetric
nuclei when the lowest-order
post-Newtonian (PN) corrections are included in the equations of motion.
The 1PN accelerations imply an orbit-averaged rate of of periapsis advance
\begin{equation}
\left(\frac{d\omega}{dt}\right)_\mathrm{GR} = \frac{2\pi}{\trad}\frac{3G\Mbh}{c^2 a \calR}
\end{equation}
\cite[][equation~(4.205)]{DEGN}.
The other elements of the osculating orbit exibit no secular variations at this PN order,
and we ignore rotation of the \sbh.
Expressed in terms of the dimensionless time variable $\tau\equiv 2\pi t/\tprec$
defined in equation~(\ref{eq_motion}), the relativistic precession rate becomes
\begin{equation}  \label{eq_kappa}
\left(\frac{d\omega}{d\tau}\right)_\mathrm{GR} = \frac{\kappa}{\calR},\ \ \ \ 
\kappa\equiv 3(2-\gamma)\frac{\rg}{a} \frac{\Mbh}{M_\star(a)} = 3(2-\gamma)(3-\gamma)\frac{\rg}{a}
\frac{\Mbh}{4\pi a^3\rho_\star(a)} 
\end{equation}
which can be added to the orbit-averaged equation of motion for $\omega$,
equation~(\ref{eq_motion}).
Setting $\gamma=1$ ($Q=3/2$) in the Hamiltonian (\ref{eq_Happrox}),
the two nontrivial equations of motion become
\begin{subequations}\label{eq:dwdldtGR}
\begin{eqnarray}
\frac{d\ell}{d\tau} &=& -\frac32\epsilon_p (1-\ell^2) \sin^2i\sin(2\omega),
\label{eq_dwdldtGRa} \\
\frac{d\omega}{d\tau} &=& \frac{\kappa}{\ell^2} - \left(1+\frac{2\epsilon_p}{3}\right)\ell
+ \epsilon_p\left[\frac{3}{\ell}\left(\cos^2i - \ell^2\right)\sin^2\omega\right].
\label{eq_dwdldtGRb}
\end{eqnarray}
\end{subequations}
If there is a fixed point, its angular momentum can be found by setting
$\dot\omega=0$ when $\omega=\pi/2$, or
\begin{equation}\label{eq_GRfp}
0=\left(1+\frac53\epsilon_p\right)\ell^4 - 3\epsilon_p\ell_z^2 - \kappa\ell.
\end{equation}
As $\kappa$ is increased from zero, the value of $\ell$ at the fixed point increases, 
reaching $\ell=1$ when
\begin{equation}
\kappa \equiv \kappa_2 = 1 + \frac53\epsilon_p - 3\epsilon_p\ell_z^2 .
\end{equation}
For $\kappa>\kappa_2$ all orbits are tubes.
For $\kappa_1\le\kappa\le\kappa_2$, where
\begin{equation}
\kappa_1 = 1-\frac{4\epsilon_p}{3},
\end{equation}
there is one family of tubes, passing below the fixed point in the $(\calR,\omega)$ plane,
and a family of saucer-like orbits that librate around the fixed point.
For $\kappa<\kappa_1$, the separatrix encloses the fixed point and there are two
families of tubes, at low and high angular momenta.
In the latter case, the angular momentum associated with the separatrix at $\omega=0$
can be found by
setting $\dot\omega=0$ in equation~(\ref{eq_dwdldtGRb}):
\begin{equation}
\calR_\mathrm{sep,GR} \equiv \ell_\mathrm{sep,GR}^2 = \left(\frac{3\kappa}{3-4\epsilon_p}\right)^{2/3}. 
\end{equation}
\begin{figure*}
$$\includegraphics[width=0.45\textwidth]{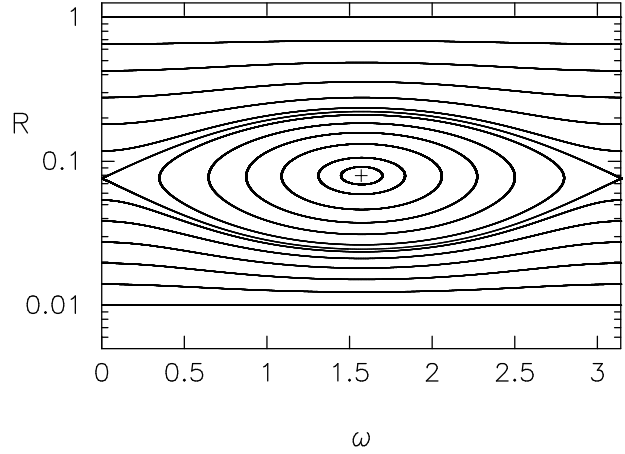}$$
\caption{
Solutions to the equations of motion (\ref{eq:dwdldtGR}) for $\calR_z=0.01$,
$\kappa=0.02$ and $\epsilon_p=0.035$.
The fixed point, equation (\ref{eq_GRfp}), is shown by the cross.
There are two familes of tube orbits: tube orbits above the fixed point, which
are present even when $\kappa=0$ (no GR); and tube orbits below the fixed point,
having sufficiently small $\ell$ that relativistic precession quenches the effects of
torques due to the flattened potential.
Equation (\ref{eq_ellminGR}) predicts $\calR\approx 1.4\times 10^{-2}$ for the
minimum $\calR$ reached by saucers; the actual value is $\sim 2.4\times 10^{-2}$.
} \label{fig_GRorbits}
\end{figure*}
Since $\kappa_1\lesssim 1\lesssim \kappa_2$,
saucer-like orbits are present only when
\begin{equation}
\frac{a}{\rg} \gtrsim 3\frac{\Mbh}{M_\star(a)}.
\end{equation}

Even when saucers are present, their angular momentum variations are limited
by the relativistic term.
A rough lower limit on the attainable angular momentum for saucers can be derived
by equating the change in $\ell$ due to the torques over one GR precessional period
to $\ell$:
\begin{equation}
\ell = \left|\frac{d\ell}{d\tau}\right| \times 
\left|\frac{1}{\pi}\frac{d\omega}{d\tau}\right|_\mathrm{GR}^{-1}
\end{equation}
which yields
\begin{equation}\label{eq_ellminGR}
\ell_\mathrm{min}\approx \frac{2}{3\pi}\frac{\kappa}{\epsilon_p}.
\end{equation}

Figure~\ref{fig_GRorbits} shows numerical solutions of the equations of motion
(\ref{eq:dwdldtGR}) for $\calR_z=0.01$ and $\kappa=0.02$; the nuclear
flattening parameter has the same value as in Figure 2.
The neglect of GR precession on the evolution of saucer orbits is justified if 
\begin{equation}  \label{eq_GRcondition}
\ell_\mathrm{min} \lesssim \sqrt{\calRcapt} \approx \sqrt{\frac{2 r_\mathrm{lc}}{a}}  
  \gtrsim 4\sqrt{\frac{\rg}{a}} \;,
\end{equation}
where the latter inequality expresses the fact that $r_\mathrm{lc}=8\rg$ for direct captures 
and larger than that for tidal disruptions. 
Requiring that $\ell_\mathrm{min} \lesssim 4\sqrt{\rg/a}$ and using equations (\ref{eq_rinfl},
\ref{eq_kappa}, \ref{eq_ellminGR}), we obtain
\begin{subequations}  \label{eq_GRcondition2}
\begin{eqnarray}
\frac{1}{2\pi\epsilon_p} &\lesssim& \frac{M_\star(a)}{\Mbh}\sqrt{\frac{a}{\rg}} =
  2\left(\frac{a}{\rh}\right)^{\frac{7}{2}-\gamma} \sqrt{\frac{\rh}{\rg}} \\
\frac{a}{\rh} &\gtrsim& \left( \frac{1}{2\pi\epsilon_p} \frac{\sigma}{c} \right)^{\frac{2}{7-2\gamma}} ,
\end{eqnarray}
\end{subequations}
where the second line approximates $\rh\approx G\Mbh/\sigma^2$.
This is essentially the same condition that was obtained in \citet{MerrittVasiliev2011} 
for capture of pyramid orbits.
Since the quantity in the brackets is likely to be small (unless $\epsilon_p$ is tiny), 
and because most of the flux into the \sbh\ comes from orbits with
$a\approx \rh$, equation (\ref{eq_GRcondition2}) suggests that relativity
is not likely to be important for the total capture rate.

\section{Orbit-averaged Hamiltonian}  \label{sec_appendix_Hamiltonian}

Here we give the exact expression for the orbit-averaged Hamiltonian 
corresponding to the potential of equation~(\ref{eq_modelrhophi}).

It is  convenient to express $r$ and $z$ in terms of 
eccentric anomaly $\eta$ rather than mean anomaly $w$ \citep[e.g.][eq.1]{SambhusSridhar2000}: 
\begin{equation}  \nonumber
r = a (1-e\cos\eta) \;,\;\; 
z = a \sqrt{1-\frac{\ell_z^2}{\ell^2}} 
  \left[ \sin\omega (\cos\eta-e) + \sqrt{1-e^2} \cos\omega\sin\eta \right] \;,\;\;
e \equiv \sqrt{1-\ell^2} .
\end{equation}
Then
\begin{equation}  \nonumber
\overline \Phi_\star \equiv \frac{1}{2\pi} \int_0^{2\pi} \Phi_\star(\mathbf{r})\,dw 
  = \frac{1}{2\pi} \int_0^{2\pi} \Phi_\star(\mathbf{r})\,d\eta\,(1-e\cos\eta) 
  = \Phi_0 \left(\frac{a}{r_0} \right)^{2-\gamma} \, \tilde H(\ell, \ell_z, \omega) ,
\end{equation}
where $\tilde H$ is expressed in terms of Gauss' hypergeometric function:
\begin{equation}  \nonumber
\tilde H = \left(1-\frac{\epsilonp}3\right) 
  {}_2F_1 \left( \frac{\gamma-2}2,\frac{\gamma-3}2;\, 1;\, 1-\ell^2 \right) 
  + \epsilonp \left\{ \frac{(4-\gamma)(5-\gamma)}{8} \;
    {}_2F_1 \left( \frac{\gamma}2,\frac{\gamma-1}2;\, 3;\, 1-\ell^2 \right)
    (1-\ell^2) \sin^2\omega  \right.
\end{equation}
\begin{equation}
  + \left. \left[ {}_2F_1 \left( \frac{\gamma}2,\frac{\gamma-1}2;\, 3;\, 1-\ell^2 \right) + 
    \frac{\gamma(\gamma-1)}{24}(1-\ell^2)\; 
    {}_2F_1 \left( \frac{\gamma+2}2,\frac{\gamma+1}2;\, 4;\, 1-\ell^2 \right)
    \right] \frac{\ell^2}{2} \right\}
    \left(1-\frac{\ell_z^2}{\ell^2}\right) .
\end{equation}

\section{Relation to the Lidov-Kozai problem}  \label{sec_appendix_kozai}

Motion in the hierarchical three-body problem is often derived from a doubly-averaged
Hamiltonian after expressing the equations of motion in Jacobi coordinates and
retaining only the lowest-order (quadrupole) term in the perturbation potential.
The ``inner restricted problem''  \citep{Kozai1962, Lidov1962} assumes furthermore 
that the test mass orbits well inside the perturber mass.
The averaged Hamiltonian describing the test particle, 
with an appropriately chosen unit of time, is
\begin{equation}  \label{eq_Ham_Kozai}
H_\mathrm{K} = -\frac{5}{6} + \frac{\calR}{2} + 
  \left(1-\frac{\calR_z}{\calR}\right) \left[\frac52(1-\calR)\sin^2\omega + \frac{\calR}{2} \right] 
\end{equation}
(Merritt 2013, equation 4.315)
which may be obtained from equation~\ref{eq_Hamiltonian_approx}
by setting $\CoefN=5/2$ and eliminating unity in the first bracket of the first term 
(equivalent to taking the limit $\epsilonp\to\infty$, eliminating terms that do not contain 
$\epsilonp$ and normalizing the unit of time to $\epsilonp$).
Equation~(\ref{eq_CalHb}) then yields a value of $5/3$ for the parameter $\calRsep$. 
As in the axisymmetric problem, motion in the Lidov-Kozai problem also exhibits two regimes: 
circulation in $\omega$ (corresponding to tube orbits) 
or libration about $\omega=\pi/2$ (analogous to saucers), the latter appearing for $\calR_z<3/5$.
The equation of motion for $\calR$ (\ref{eq_dcalRdtau}) is
\begin{equation}
\frac{\d\calR}{d\tau} = -2\sqrt{6}\,\sqrt{(\calR_1-\calR)(\calR-\calR_2)(\calR_3-\calR)} \;,
\end{equation}
with the same relation between $\calR_1,\calR_2,\calR_3$ as in 
equation~(\ref{eq_Rminmax}):
\begin{subequations}
\begin{eqnarray}
\calR_{1,2} &\equiv& \calR_\star \pm \sqrt{\calR_\star^2 - (5/3)\calR_z} \\
\calR_3     &\equiv& H_\mathrm{K}+5/6+\calR_z/2 \\
\calR_\star &\equiv& (5+5\calR_z-2\calR_3)/6 .
\end{eqnarray}
\end{subequations}
The case of circulation corresponds to $\calR_2 \le \calR \le \calR_3 \le \calR_1$, 
and libration to $\calR_2 \le \calR \le \calR_1 \le \calR_3$; 
in the latter case the relation between minimum and maximum values of $\calR$ 
is the same as in equation~(\ref{eq_Rminmax_saucer}), from which it follows that the 
librating regime exists for $\calR_z<1/\calRsep=3/5$.
The separatrix between ``tubes'' and ``saucers'' is at $H_\mathrm{K}=1/6-\calR_z/2$
and the fixed-point saucer has $H_\mathrm{K}=5/3+2\calR_z-\sqrt{15\calR_z}$.
For comparison, in the oblate axisymmetric potential considered throughout this paper, 
the first regime (tube orbits) has $\calR_2 \le \calR_3 \le \calR \le \calR_1$ 
and the second (saucers) has $\calR_3 \le \calR_2 \le \calR \le \calR_1$.

\section{Saucer orbits beyond the sphere of influence}   \label{sec_appendix_saucers}

Beyond the influence sphere, orbits similar to the saucers can still exist in axisymmetric
potentials \citep{Richstone1982, LeesSchwarzschild1992, Evans1994}, 
but they are not describable in terms of osculating Keplerian elements.
Typically such orbits are described as tube orbits that lie close to a resonance
between the radial and vertical motions.
To make the correspondence with our work more clear, 
we recast the fixed-point saucer orbit
near a \sbh\ in terms of the Cartesian variables ($R, z$), i.e. 
cylindrical coordinates in the meridional plane.
Setting $\omega=\pi/2$ in  equation (A1) yields for the fixed-point orbit that generates 
the saucers
\begin{equation}\label{eq_meridional_fp}
\frac{R^2}{a^2} = \left(1-e^2\right)^2 - 
\frac{2e\left(1-e^2\right)}{\sin i} \frac{z}{a} -
\left(1-\frac{e^2}{\sin^2i}\right)\frac{z^2}{a^2}
\end{equation}
where it is understood that $e$ and $\cos i$ have their fixed-point values:
\begin{equation}
e^2= 1 - \sqrt{\calRsep\calR_z}, \ \
\cos^2 i =\sqrt{\frac{\calR_z}{\calRsep}}. \nonumber
\end{equation}
Equation (\ref{eq_meridional_fp})  is a hyperbola in the ($R,z$) plane, between
the points
\begin{eqnarray}
R_\mathrm{max} &=& (1+e)a\cos i, \ \ z_\mathrm{max} = (1+e)a\sin i, \nonumber \\
R_\mathrm{min} &=& (1-e)a\cos i, \ \ z_\mathrm{max} = (1-e)a\sin i. \nonumber
\end{eqnarray}
The curve crosses the equatorial plane at $R = a\sqrt{\calRsep\calR_z}=a(1-e^2)$, the semi-latus rectum.

When the fixed-point saucer orbit first appears, at $\calR_z=\calRsep$, it lies 
in the equatorial plane ($\cos i = 1, z_\mathrm{max} = 0$) and has zero 
vertical thickness.
For $\calR_z$ values smaller than this maximum, the trajectory (\ref{eq_meridional_fp}) 
can be interpreted as a $1:1$ resonance between motions in the $R$ and $z$ directions.

Described in this way,  saucer orbits near a \sbh\ are seen to have very similar properties to 
orbits described by other authors in more general axisymmetric potentials.
\citet{LeesSchwarzschild1992} studied orbits in scale-free axisymmetric
models with logarithmic potentials, $\rho \sim r^{-2}$ and no central \sbh.
For models with density axis ratio $0.265$, they found that saucers first appear
at $\calR_z\approx 0.48$; for smaller $\calR_z$ the fixed-point orbit
(which they called a ``reflected banana'') traces a path in the ($R,z$) plane
similar to equation (\ref{eq_meridional_fp}). 
They noted that motion near the fixed-point orbit is regular, i.e. non-chaotic.
Similar orbits were described by \citet{Richstone1982} (who called them ``pipe orbits'')
and \citet{Evans1994} in surveys 
of orbits in other scale-free families of oblate models.

\section{Local diffusion coefficients}  \label{sec_appendix_difcoefs}

Here we present the local (position-dependent) diffusion coefficients appearing 
in equation~(\ref{eq_FPlocal}), expressed in the spherical coordinates 
$(r, \theta, \phi)$ and generalized velocities 
$(\calE=-\Phi(r)-v^2/2$, $\calR=\Lsqinv L^2$, $\calR_z=\Lsqinv L_z^2)$.
We denote $\Lsqinv \equiv \Lcirc^{-2}(E)$, $\Lsqinv'=d\Lsqinv/dE = -d\Lsqinv/d\calE$, 
$\calQ = 1 + (v^2/2)(\kappa^\prime/\kappa)$.

\begin{eqnarray}  \label{eq_dif_coefs_local}
\langle \Delta \calE \rangle  &=&  -v\dvpar - \frac{1}{2}\dvsqpar - \frac{1}{2}\dvsqper ,\\
\langle \Delta \calR \rangle  &=& 2\calR\calQ \frac{\dvpar}{v} + 
  \left(\frac{v^4}{2}\Lpp + 5\calQ - 4\right) \calR \frac{\dvsqpar}{v^2} + 
  \left(\calQ - \frac{3}{2} + \frac{v^2}{v^2-v_r^2}\right) \calR \frac{\dvsqper}{v^2} , \nonumber\\
\langle \Delta \calR_z \rangle  &=& 2\calR_z\calQ \frac{\dvpar}{v} + 
  \left(\frac{v^4}{2}\Lpp + 5\calQ - 4\right) \calR_z \frac{\dvsqpar}{v^2} + 
  \left(\frac{\calR\,\sin^2\theta}{2\calR_z}\, \frac{v^2-v_\phi^2}{v^2-v_r^2} + \calQ-1\right)
  \calR_z\,\frac{\dvsqper}{v^2}  , \nonumber\\
\langle(\Delta\calE)^2\rangle  &=& v^2 \dvsqpar , \nonumber\\
\langle(\Delta\calR)^2 \rangle  &=& 4\calR^2\calQ^2\frac{\dvsqpar}{v^2} + 
  2\calR^2 \frac{v_r^2}{v^2-v_r^2}\, \frac{\dvsqper}{v^2} , \nonumber\\
\langle(\Delta\calR_z)^2 \rangle  &=& 4\calR_z^2\calQ^2\frac{\dvsqpar}{v^2} + 
  2\calR \calR_z\,\sin^2\theta\, \frac{v^2-v_\phi^2}{v^2-v_r^2}\, \frac{\dvsqper}{v^2} , \nonumber\\
\langle \Delta \calE \Delta \calR \rangle  &=& -2\calR\calQ \dvsqpar , \nonumber\\
\langle \Delta \calE \Delta \calR_z\rangle  &=& -2\calR_z\calQ \dvsqpar , \nonumber\\
\langle \Delta \calR \Delta \calR_z\rangle  &=& 4\calR\calR_z \calQ^2 \frac{\dvsqpar}{v^2} + 
  2\calR\calR_z \frac{v_r^2}{v^2-v_r^2}\,\frac{\dvsqper}{v^2} . \nonumber
\end{eqnarray}
These coefficients have been expressed in terms of velocity diffusion coefficients for 
$\{v_\|, v_\bot\}$ via integrals of the distribution function $f(\calE_f)$ describing the
field stars (of mass $m_\star$)
using the relative potential $\Psi\equiv -\Phi$:
\begin{eqnarray}  \label{eq_veldiffcoef}
v\dvpar &=& \textstyle -\left(1+\frac{m}{m_\star}\right) I_{1/2} ,\\
\dvsqpar&=& \textstyle \frac{2}{3} \left(I_0 + I_{3/2}\right) , \nonumber\\
\dvsqper&=& \textstyle \frac{2}{3} \left(2 I_0 + 3 I_{1/2} - I_{3/2}\right) ,\nonumber
\end{eqnarray}
where
\begin{eqnarray}
I_0     &\equiv& 16\pi^2G^2m_\star^2\ln\Lambda \int_0^\calE d\calE_f\,f(\calE_f) , \\
I_{n/2} &\equiv& 16\pi^2G^2m_\star^2\ln\Lambda \int_\calE^{\Psi(r)} d\calE_f\,f(\calE_f)
  \left(\frac{\Psi-\calE_f}{\Psi-\calE}\right)^{n/2} . \nonumber
\end{eqnarray}
When making comparisons with the relaxation rates measured from $N$-body 
simulations we averaged the diffusion coefficients over the subspace $\calE=\const$:
\begin{eqnarray}  \label{eq_dif_coefs_avg_E}
\langle(\Delta\calE)^2\rangle_\mathrm{av} &=& p^{-1}(\calE) \int_0^{r_\mathrm{max}(\calE)} dr\; r^2v\; 
  v^2\dvsqpar \\
\langle(\Delta\calR)^2\rangle_\mathrm{av} &=& p^{-1}(\calE) \int_0^{r_\mathrm{max}(\calE)} dr\; r^2v\; 
  \frac{4\, r^4v^2}{15\, \Lcirc^2} \left(8Q^2\dvsqpar + \dvsqper\right) , \nonumber\\
p(\calE) &\equiv& \int_0^{r_\mathrm{max}(\calE)} dr\; r^2v \;,\quad 
v\equiv\sqrt{2(\Psi(r)-\calE)},\;\;r_\mathrm{max}\equiv \Psi^{-1}(\calE) . \nonumber
\end{eqnarray}

\section{Coordinates in 2D Fokker-Planck equation}  \label{sec_appendix_coords}

Although we discuss the two-dimensional Fokker-Planck equation in the $\calH-\calR_z$ plane 
throughout the paper, it is more convenient to obtain the  numerical solution using a different 
set of coordinates, for which we 
have implemented two variants: $(\nu, \mu)$ and $(\nu, \xi)$, where
\begin{equation}
\nu \equiv \calH+\calR_z \;;\quad
\mu \equiv \frac{1}{2}\left( b-\sqrt{b^2-\frac{4\calR_z}{\calRsep}}\right) , \quad  
  b \equiv \frac{\calH+\calR_z}{\calRsep} + 1-\calH \;;\quad
\xi \equiv \frac{\calR_z}{\calH-\calH_0} \;,\quad   \calH_0 \equiv -\frac{\calRsep-\calRcapt}{1-\calRsep} .
\end{equation}
The lines of constant $\nu$ are diagonal lines in the $\calH-\calR_z$ plane, parallel 
to the capture boundary in the tube ($\calH>0$) region, and lines of constant $\mu$ or $\xi$ 
are also straight lines in that plane,
designed in such a way that the capture boundary in the saucer region ($\calH>0$) has 
$\mu=\mathrm{const}=\calRcapt/\calRsep$ or $\xi=\mathrm{const}=-\calRcapt/\calH_0$. This facilitates 
setting boundary conditions on these capture boundaries which are parallel to coordinate axes. 
We used both coordinate sets to cross-check the results.

\end{document}